\documentclass[useAMS]{mn2e}
\usepackage{graphicx}
\usepackage{amssymb}
\usepackage{amsmath}
\usepackage{color}

\newcommand\be{\begin{equation}}
\newcommand\ee{\end{equation}}

\newcommand\ba{\begin{eqnarray}}
\newcommand\ea{\end{eqnarray}}

\title{Tidal Truncation of Inclined Circumstellar and Circumbinary Discs in Young Stellar Binaries}

\author [R. Miranda and D. Lai]
        {Ryan Miranda and Dong Lai \\
        Center for Space Research, Department of Astronomy, Cornell University, Ithaca, NY 14853, USA}

\begin{document}

\maketitle

\begin{abstract}
Recent observations have shown that circumstellar and circumbinary discs in young stellar binaries are often misaligned with respect to the binary orbital plane. We analyze the tidal truncation of such misaligned discs due to torques applied to the disc at the Lindblad resonances from the tidal forcings of the binary. We consider eccentric binaries with arbitrary binary-disc inclination angles. We determine the dependence of the tidal forcing strengths on the binary parameters and show that they are complicated non-monotonic functions of eccentricity and inclination. We adopt a truncation criterion determined by the balance between resonant torque and viscous torque, and use it to calculate the outer radii of circumstellar discs and the inner radii of circumbinary discs. Misaligned circumstellar discs have systematically larger outer radii than aligned discs, and are likely to fill their Roche lobes if inclined by more than $45^\circ - 90^\circ$, depending on the binary mass ratio and disc viscosity parameter. Misaligned circumbinary discs generally have smaller inner radii than aligned discs, but the details depend sensitively on the binary and disc parameters.
\end{abstract}


\section{Introduction}

Protoplanetary discs are often found in binary systems. In general, because of the complex (and often turbulent) formation processes of stars, binaries, and discs (e.g., Bate, Bonnell \& Bromm 2003; McKee \& Ostriker 2007; Klessen 2011; Fielding et al. 2015), both circumstellar and circumbinary discs are likely formed with inclined orientations with respect to their binary orbital planes. Indeed, a number of binary young stellar objects (YSOs) are observed to contain circumstellar discs that are misaligned with the binary plane (e.g., Stapelfeldt et al. 1998, 2003; Neuh\"{a}user et al. 2009). The orientation of jets in several unresolved YSOs or pre-main-sequence binaries also suggest misalignment (e.g., Davis, Mundt \& Eisl\"{o}ffel 1994; Roccatagliata et al. 2011). Recently, Jensen \& Akeson (2014) showed evidence that at least one of the circumstellar discs in the young binary system HK Tau is inclined with respect to the binary plane by at least $30^\circ$. Imaging of circumbinary debris discs indicates that while some are aligned with the binary plane, others can have significant misalignment (such as 99 Herculis, which is misaligned by at least $30^\circ$; Kennedy et al. 2012a, 2012b). The pre-main-sequence binary KH 15D is surrounded by a precessing circumbinary disc inclined with respect to the binary plane by $10-20^\circ$ (Chiang \& Murray-Clay 2004; Winn et al. 2004; Capelo et al. 2012), and in the FS Tau system, circumstellar discs appear to be misaligned with a circumbinary disc (Hioki et al. 2011). Although circumbinary discs may align with the binary plane on relatively short timescales, circumstellar discs are expected to maintain misalignment over timescales comparable to their lifetimes (Foucart \& Lai 2014).

In recent years, many exoplanets have been found in binaries, including both S-type planets (orbiting a single member of the binary) using radial velocity measurements (e.g., Hatzes et al. 2003; Chauvin et al. 2011; Dumusque et al. 2012), and P-type (circumbinary) planets using the transit method (e.g., Doyle et al. 2011; Orosz et al. 2012; Welsh et al. 2012, 2014; Kostov et al. 2014). While the orbits of most of the circumbinary planets are consistent with alignment with the binary orbit, Kepler-413b appears to be slightly misaligned (by about $2.5^\circ$), possibly as a result of forming in a misaligned circumbinary disc (Kostov et al. 2014).

The sizes of the protoplanetary discs (the outer radius of a circumstellar disc and the inner radius of a circumbinary disc) in which planets in binaries form are affected by gravitational interactions between the disc and binary. The size of the disc is of interest because it places restrictions on the process of planet formation. For example, the process of giant planet formation may be jeopardized in circumstellar discs with significantly truncated outer radii (Jang-Condell 2015).

The size of a pressureless circumstellar disc is determined by the ``static'', non-resonant tidal force from the binary companion. In particular, the disc size is determined by the location at which orbits around the primary star begin to intersect one another, in the context of the circular restricted three-body problem (Paczy\'{n}ski 1977; Paczy\'{n}ski \& Rudak 1980). In this framework, the outer edge of a circumstellar disc cannot extend past the $2$:$1$ commensurability (i.e., the period ratio of the binary and a test mass in the disc must exceed two), except when the binary mass ratio is extreme. Pichardo, Sparke \& Aguilar (2005) extended this analysis to eccentric coplanar binaries, and found that the disc must always be smaller than $73\%$ of the average Roche lobe radius. However, Papaloizou \& Pringle (1977) demonstrated that even a very small viscosity can prevent orbit crossings and allow the disc to be larger than the radius determined by non-intersecting orbits.

In addition to static tidal interactions, gas discs (both circumstellar and circumbinary) in binaries are also subject to resonant tidal forcings. The motion of the binary excites spiral density waves at Lindblad resonances, where the natural epicyclic frequency $\kappa$ of the disc is commensurate with the binary orbital frequency $\Omega_\mathrm{B}$. As a result, a torque is exerted on the disc at each resonance (Goldreich \& Tremaine 1979, 1980). Artymowicz \& Lubow (1994; hereafter AL94) applied the formalism of Goldreich \& Tremaine to determine the sizes of circumstellar and circumbinary discs in coplanar binaries. In their approach, the gravitational potential of an eccentric binary is decomposed into many Fourier components, each applying a torque on the disc at Lindblad resonances. The possibility of gap opening and disc truncation by each potential component is determined by the balance between the Lindblad torque and the viscous torque. The main goal of this paper is to generalize the results of AL94 to circumstellar and circumbinary discs in binaries with arbitrary mutual inclinations.

The behavior of Lindblad torques in misaligned discs has been explored in several recent works. Lubow, Martin \& Nixon (2015) investigated the torques experienced by a circumstellar disc in an inclined circular binary, which is a special case of the general binary eccentricities studied in this paper. Nixon \& Lubow (2015) demonstrated that an eccentric binary can exert non-zero Lindblad torques on a retrograde circumbinary disc, which are significantly weaker than those applied to a prograde disc, but can nonetheless be responsible for clearing a cavity around the binary. This is a special case of the arbitrary inclination framework used in this paper, although they did not make the assumption of a Keplerian disc. We do not specifically address retrograde discs in this paper.

The most important technical aspect of our work is the computation of the Fourier components of the binary potential for arbitrary inclinations. We represent the potential semi-analytically as a power series in eccentricity (as in the appendix of AL94), evaluated to order $e^{10}$. This is accomplished through the use of Wigner matrices to account for arbitrary inclination between the disc and binary planes, and through an expansion of the binary orbit using the Hansen coefficients of celestial mechanics. We note that Nixon \& Lubow (2015) utilized an exact, non-series approach for computing the gravitational potential for the special case of retrograde circumbinary discs. This method amounts to numerically evaluating the Hansen integrals.

This paper is organized as follows. In Section \ref{sec:potential} we decompose the gravitational potential of a misaligned binary into Fourier components with various forcing frequencies. In Section \ref{sec:torque} we review how these potential components exert Lindblad torques on a disc. In Section \ref{sec:cs_disc} we apply this theory to the truncation of circumstellar discs, determining the size of the disc outer edge. We then determine the inner radii of circumbinary discs in Section \ref{sec:cb_disc}. We address the validity of our Keplerian disc approximation in Section \ref{sec:non_kep}. We summarize and discuss our results in Section \ref{sec:discussion}. 

\section{Potential Components}
\label{sec:potential}

We consider a binary consisting of masses $M_1$ and $M_2$ with total mass $M_\mathrm{tot} = M_1 + M_2$, mass ratio $q = M_2/M_\mathrm{tot}$, semi-major axis $a$ and eccentricity $e$. The orbital frequency is $\Omega_\mathrm{B} = \left(GM_\mathrm{tot}/a^3\right)^{1/2}$. The orbit is inclined relative to a reference plane, taken to be the plane of a thin circumstellar or circumbinary disc, by an angle $i$. We will express the disturbing potential (per unit mass) acting on the disc in the form
\be
\begin{aligned}
\Phi &= \sum_{m,\mu,n} \Phi_{m,\mu,n}\left(r\right) \cos\left[m\phi-\left(\mu+n\right)\Omega_\mathrm{B}t\right] \\
&= \sum_{m,N} \Phi_{m,N}\left(r\right) \cos\left(m\phi-N\Omega_\mathrm{B}t\right),
\end{aligned}
\ee
where $\left(r,\phi\right)$ specifies the radial and azimuthal position of the disc particle, $m$ is the azimuthal number in the disc plane, $\mu$ is the azimuthal number in the binary plane, $N$ is a time harmonic number, and $n$ is related to the eccentricity dependence of each potential component (see later in this section). In our formulation, $m > 0$, but $\mu$ and $n$ can be positive, negative, or zero. Since different $(\mu,n)$-components with the same $\phi$ and $t$ dependence are not physically distinct, we have defined
\be
\Phi_{m,N}\left(r\right) = \sum_{\mu,n} \delta_{\mu+n,N} \Phi_{m,\mu,n}\left(r\right),
\ee
which measures the total strength of the potential component having the azimuthal number $m$ and rotating with the pattern frequency
\be
\omega_\mathrm{P} = \frac{N}{m}\Omega_\mathrm{B}.
\ee
Each component, denoted by $\left(m,N\right)$, excites density waves at resonant locations in the disc, which give rise to the torques, $T_{m,N}$. We use different approaches to compute the potential $\Phi_{m,N}$, depending on whether we are considering a circumstellar or circumbinary disc.

\subsection{Circumstellar Disc}
We work in the reference frame centered on the primary star $M_1$, so that the disc rotation rate around $M_1$ is $\Omega\left(r\right) = \left(GM_1/r^3\right)^{1/2}$. The gravitational potential felt by a disc particle due to $M_2$ is
\be
\Phi = -\frac{GM_2}{\lvert\mathbf{r}-\mathbf{r}_2\rvert} + GM_2\frac{\mathbf{r} \cdot \mathbf{r}_2}{r_{12}^3},
\ee
where $\mathbf{r}$ is the particle's position vector, and $\mathbf{r}_2 = r_{12}\hat{\mathbf{r}}_2$ is the position vector of $M_2$ relative to $M_1$. The second term in the potential is the indirect part arising from the motion of $M_1$ around the center of mass of the system (e.g., Murray \& Dermott 1999). The direct term can be expanded in Legendre polynomials, leading to
\be
\label{eq:legendre}
\Phi = -GM_2 \sum_{l=2}^{\infty} \frac{r^l}{r_{12}^{l+1}} P_l\left(\hat{\mathbf{r}} \cdot \hat{\mathbf{r}}_2\right).
\ee
The $l = 0$ term is an irrelevant constant which has been dropped. The angular dependence can be separated by an expansion in spherical harmonics,
\be
\label{eq:legendre_ylm}
P_l\left(\hat{\mathbf{r}} \cdot \hat{\mathbf{r}}_2\right) = \frac{4\pi}{2l+1} \sum_{m=-l}^{l} Y^*_{l,m}\left(\theta_2,\phi_2\right) Y_{l,m}\left(\theta,\phi\right).
\ee
We define two coordinate systems. The unprimed coordinate system $\left(\theta,\phi\right)$ has its $z$-axis aligned with the disc angular momentum, $\hat{\mathbf{z}} = \hat{\mathbf{L}}_\mathrm{D}$. The primed coordinate system $\left(\theta',\phi'\right)$ has its $z$-axis aligned with the binary angular momentum, $\hat{\mathbf{z}}' = \hat{\mathbf{L}}_\mathrm{B}$. The two coordinate systems share a $y$-axis, and $\hat{\mathbf{z}} \cdot \hat{\mathbf{z}}' = \cos\left(i\right)$. The spherical harmonics in these two coordinate systems are related through the Wigner $d$ matrices,
\be
\label{eq:ylm_wigner}
Y_{l,m}\left(\theta_2,\phi_2\right) = \sum_{\mu=-l}^{l} d_{\mu,m}^{l}\left(i\right) Y_{l,\mu}\left(\theta'_2,\phi'_2\right).
\ee
Combining equations (\ref{eq:legendre}), (\ref{eq:legendre_ylm}) and (\ref{eq:ylm_wigner}), evaluating $Y_{l,m}$ in the disc plane ($\theta = \pi/2$) and $Y_{l,\mu}$ in the binary plane ($\theta' = \pi/2$), and taking only the real part, the potential (\ref{eq:legendre}) can be written as
\be
\begin{aligned}
\Phi = -&\frac{2GM_2}{a} \sum_{l=2}^{\infty} \sum_{m=1}^{l} \sum_{\mu=-l}^{l} W_{l,m} W_{l,\mu} d_{\mu,m}^{l}\left(i\right) \\
&\times \left(\frac{r}{a}\right)^l \left(\frac{r_{12}}{a}\right)^{-\left(l+1\right)} \cos\left(m\phi-\mu\phi'_2\right),
\end{aligned}
\ee
where
\be
\begin{aligned}
W_{l,m} & = \left[\frac{\left(l-m\right)!}{\left(l+m\right)!}\right]^{\frac{1}{2}} P_l^m\left(0\right) \\
& = \left(-1\right)^{\frac{l+m}{2}} \left[\left(l-m\right)!\left(l+m\right)!\right]^{\frac{1}{2}} \\
&\times \left[2^l \left(\frac{l-m}{2}\right)! \left(\frac{l+m}{2}\right)!\right]^{-1},
\end{aligned}
\ee
with the factor $(-1)^{(l+m)/2}$ taken to be zero if $l+m$ is odd, and similarly for $W_{l,\mu}$. The product $W_{l,m}W_{l,\mu}$ is zero unless $l$, $m$ and $\mu$ are all even or all odd. Notice that we have taken twice the sum over only positive values of $m$, since the $\left(-m,-\mu\right)$ and $\left(m,\mu\right)$ terms are identical (those with positive or negative $\mu$ are still distinct). In this form, the expression for the potential is exact, given an explicit forms of $r_{12}/a$ and $\phi'_2$, the radial coordinate and true anomaly of $M_2$ in the binary plane. As a final step, we use their elliptic expansions to write (see Appendix \ref{sec:elliptic})
\be
\begin{aligned}
\label{eq:cs_expansion}
&\left(\frac{r_{12}}{a}\right)^{-\left(l+1\right)} \cos\left(m\phi-\mu \phi'_2\right) = \\ 
&\sum_{n=-\infty}^\infty C_{l,\mu,n}^{\mathrm{CS}} \cos\left[m\phi-\left(\mu+n\right)\Omega_\mathrm{B}t\right].
\end{aligned}
\ee
Each coefficient $C^{\mathrm{CS}}_{l,\mu,n}$ is a series in powers of $e$, with the leading term proportional to $e^{\lvert n\rvert}$. The main approximation we make is to truncate these coefficients at a finite order in eccentricity ($e^{10}$).  The final expression for the potential strengths is
\be
\Phi_{m,\mu,n} = -\frac{2GM_2}{a} \sum_{l=l_\mathrm{min}}^\infty W_{l,m} W_{l,\mu} C_{l,\mu,n}^{\mathrm{CS}} d_{\mu,m}^l\left(i\right) \left(\frac{r}{a}\right)^l,
\ee
where $l_\mathrm{min} = \max\left(m,\lvert \mu \rvert,2\right)$. 

\subsection{Circumbinary Disc}
For circumbinary discs, we work in the reference frame centered on the center of mass of the binary. In this frame, the orbital frequency of a particle in the disc is $\Omega\left(r\right) = \left(GM_\mathrm{tot}/r^3\right)^{1/2}$. The disturbing potential can then be expressed as
\be
\label{eq:cb_potential}
\Phi = -\sum_{l=2}^\infty \frac{GM_l}{a} \left(\frac{r_{12}}{a}\right)^l \left(\frac{r}{a}\right)^{-\left(l+1\right)} P_l\left(\hat{\mathbf{r}} \cdot \hat{\mathbf{r}}_2\right),
\ee
where $r_{12}$ is the separation between $M_1$ and $M_2$, and
\be
M_l = \left[q\left(1-q\right)^l + \left(-1\right)^l \left(1-q\right)q^l\right]M_\mathrm{tot}
\ee
(e.g., Ford, Kozinsky \& Rasio 2000; Harrington 1968). In general, odd-$l$ components are weaker than even-$l$ components for similar mass binaries, and for equal mass binaries ($q = 1/2$), $M_l$ is identically zero for odd $l$. Again writing the Legendre polynomials in terms of spherical harmonics in the binary and disc frames using the Wigner functions [equations (\ref{eq:legendre_ylm}) and (\ref{eq:ylm_wigner})], we have
\be
\begin{aligned}
\Phi = -2&\sum_{l=2}^\infty \frac{GM_l}{a} \sum_{m=1}^l \sum_{\mu=-l}^l W_{l,m} W_{l,\mu} d_{\mu,m}^l\left(i\right) \\
& \times \left(\frac{r}{a}\right)^{-\left(l+1\right)} \left(\frac{r_{12}}{a}\right)^l \cos\left(m\phi-\mu\phi'_2\right).
\end{aligned}
\ee
We expand the time-dependent orbital coordinates of the binary in a manner analogous to the circumstellar disc case,
\be
\label{eq:cb_expansion}
\begin{aligned}
&\left(\frac{r_{12}}{a}\right)^l \cos\left(m\phi-\mu \phi'_2\right) = \\
&\sum_{n=-\infty}^\infty C^\mathrm{CB}_{l,\mu,n}\cos\left[m\phi-\left(\mu+n\right)\Omega_\mathrm{B}t\right],
\end{aligned}
\ee
so the potential component is given by
\be
\Phi_{m,\mu,n} = -2 \sum_{l=l_\mathrm{min}}^\infty \frac{GM_l}{a} W_{l,m} W_{l,\mu} C_{l,\mu,n}^{\mathrm{CB}} d_{\mu,m}^l\left(i\right) \left(\frac{r}{a}\right)^{-\left(l+1\right)}.
\ee
See Appendix \ref{sec:elliptic} for the values of the $C_{l,\mu,n}^{\mathrm{CB}}$ coefficients.

\section{Lindblad Torques}
\label{sec:torque}

Each potential component $\Phi_{m,N}$, rotating with the pattern frequency $\omega_\mathrm{P} = N\Omega_\mathrm{B}/m$, excites density waves at the Lindblad resonances (LRs), where
\be
\label{eq:lr_criterion}
\omega_\mathrm{P} - \Omega\left(r\right) =  \pm \frac{\kappa\left(r\right)}{m},
\ee
where the upper (lower) sign corresponds to the outer (inner) LR. From here on we assume that the epicyclic frequency $\kappa\left(r\right)$ is equal to $\Omega\left(r\right)$ and both are proportional to $r^{-3/2}$, i.e., the disc is exactly Keplerian. The LRs correspond to locations where
\be
\label{eq:lr_omega}
\frac{\Omega\left(r\right)}{\Omega_\mathrm{B}} = \frac{N}{m \pm 1},
\ee
and are located at
\be
\frac{r_\mathrm{LR}}{a} = 
\begin{cases}
\label{eq:lr_locations}
\left[\left(m\pm1\right)/N\right]^{2/3} \left(1-q\right)^{1/3} &\mbox{circumstellar disc} \\
\left[\left(m\pm1\right)/N\right]^{2/3} &\mbox{circumbinary disc}
\end{cases}
.
\ee
The torque on the disc at a LR is (Goldreich \& Tremaine 1979)
\be
\label{eq:lr_torque}
T_{m,N}^{\mathrm{LR}} = -m\pi^2 \left[\Sigma \left(\frac{\mathrm{d}D}{\mathrm{d}\ln r}\right)^{-1}\lvert\Psi_{m,N}\rvert^2\right]_{r_\mathrm{LR}},
\ee
where $\Sigma$ is the disc surface density, $D = \kappa^2 - m^2\left(\Omega-\omega_\mathrm{P}\right)^2$, and
\be
\Psi_{m,N} = \frac{\mathrm{d} \Phi_{m,N}}{\mathrm{d}\ln r} + \frac{2 \Omega}{\Omega - \omega_\mathrm{P}} \Phi_{m,N}.
\ee
We also define $\Psi_{m,\mu,n}$, which is the same as the above expression but with $\Phi_{m,\mu,n}$ replacing $\Phi_{m,N}$. While $\Psi_{m,N}$ is the quantity which determines the torque, it is useful to discuss how various components $\Psi_{m,\mu,n}$ contribute to it, since these are physically associated with different azimuthal forcing components of the binary orbit. Note that since we assume Keplerian discs, $2\Omega/\left(\Omega-\omega_\mathrm{P}\right) = \mp 2m$ at $r = r_\mathrm{LR}$, and
\be
\left(\frac{\mathrm{d}D}{\mathrm{d}\ln r}\right)_{r_\mathrm{LR}} = \mp \frac{3N^2}{m \pm 1} \Omega^2_\mathrm{B}.
\ee
At inner Lindblad resonances (ILRs), $T_{m,N}^{\mathrm{LR}} < 0$, i.e., disc particles lose angular momentum, and at outer Lindblad resonances (OLRs), $T_{m,N}^{\mathrm{LR}} > 0$, so disc particles gain angular momentum. Torques are also applied to the disc at corotation resonances, where $\omega_\mathrm{P} - \Omega\left(r\right) = 0$, however these are not important in disc truncation. Therefore we subsequently drop the superscript ``LR'' on $T_{m,N}$, as we only consider Lindblad torques.

The viscous torque on the disc, assuming the $\alpha$-ansatz for the kinematic viscosity coefficient, $\nu = \alpha c_\mathrm{s}^2/\Omega$, is given by (e.g., Pringle 1981)
\be
T_\nu = 3\pi \alpha h^2\Sigma \Omega^2 r^4,
\ee
where $h = H/r$ is the disc aspect ratio. As in AL94, we assume a gap is opened at the $(m,N)$ LR if $|T_{m,N}| \geq |T_\nu|$. Throughout this paper, unless otherwise noted, we adopt a disc model with $h = 0.05$ and $\alpha = 0.01$.

\section{Results: Circumstellar Disc}
\label{sec:cs_disc}

\subsection{Resonances Relevant to Outer Disc Truncation}
We are interested in the largest outer radius $r_\mathrm{out}$ that a circumstellar disc can have, given the orbital parameters ($a$, $e$, and $i$) of the binary. This amounts to determining the smallest radius at which a particular resonant torque can open a gap. Therefore we should consider the inner Lindblad resonances of the potential components with the largest pattern frequencies that satisfy $|T_{m,N}| \geq |T_\nu|$, with $|T_\nu|$ evaluated at the resonance location. The pattern frequency associated with $\Phi_{m,N}$ is $\omega_\mathrm{P} = N\Omega_\mathrm{B}/m$, so we should examine components with small $m$ and positive $N$. The smallest $m$-components whose ILRs are in the disc have $m = 2$, since all $m = 1$ ILRs are formally located at the origin [see equation (\ref{eq:lr_locations})]. Therefore we focus on the $(m,N) = (2,N)$ ILRs with $N \geq 2$, which are located at the $\Omega/\Omega_\mathrm{B} = N$:$1$ commensurabilities.

\subsection{Effects of Disc Inclination}

\begin{figure}
\begin{center}
\includegraphics[width=0.45\textwidth,clip]{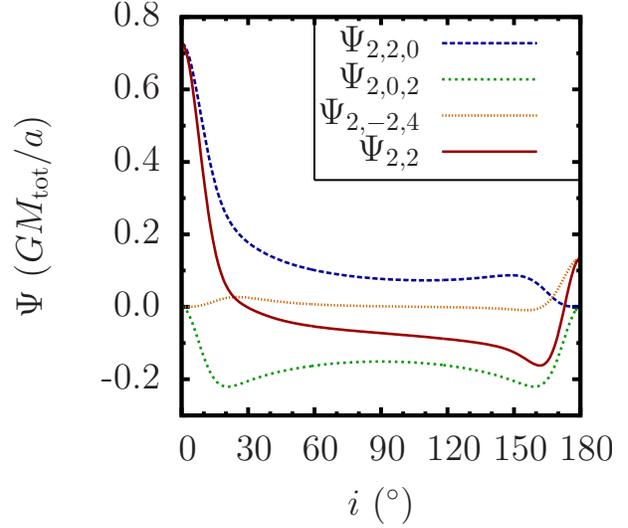}
\caption{The potential component $\Psi_{m,N} = \Psi_{2,2}$ (solid line) and the dominant $\Psi_{2,\mu,n}$ sub-components which contribute to it (dashed lines) for a circumstellar disc in an equal mass binary with $e = 0.5$.}
\label{fig:psi}
\end{center}
\end{figure}

\begin{figure*}
\begin{center}
\includegraphics[width=0.99\textwidth,clip]{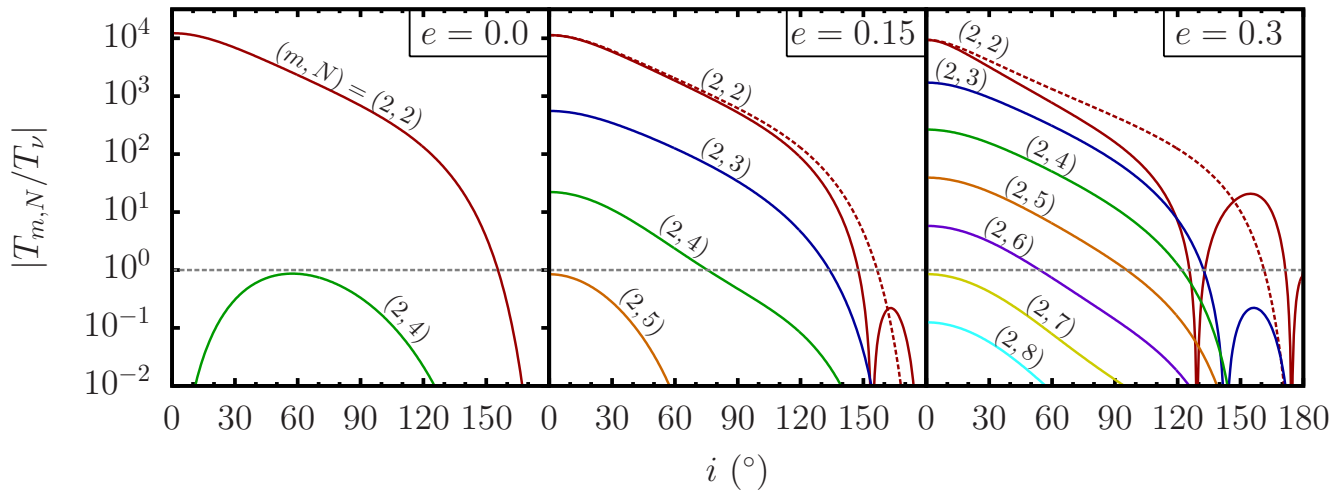}
\caption{Ratio of resonant torque to viscous torque as a function of inclination for the $(m,N) = (2,N)$ inner Lindblad resonances for a circumstellar disc in an equal-mass binary. Each panel corresponds to a different binary eccentricity. The horizontal dashed line corresponds to $|T_{m,N}| = |T_\nu|$, above which a given resonance can clear a gap. The dashed red lines (in the middle and right panels) are the $(2,2)$ torques computed using only the $\Psi_{2,2,0}$ potential component, rather than a sum of all appropriate $\Psi_{m,\mu,n}$'s.}
\label{fig:torques_cs_i}
\end{center}
\end{figure*}

An example of the different contributions to a potential component is depicted in Figure \ref{fig:psi}. Here $\Psi_{m,N} = \Psi_{2,2}$ is shown (solid line) for an equal mass binary with $e = 0.5$ as a function of disc-binary inclination $i$. The potential is the sum of all $\Psi_{2,\mu,n}$'s with $\mu + n = 2$. The dashed lines depict each of the individual components which significantly contribute to $\Psi_{2,2}$. Other components, not shown in Figure \ref{fig:psi}, such as $\Psi_{2,-4,6}$, contribute negligibly (for $e = 0.5$), but can be important at larger $e$. We include components up to $|n| = 8$ in all subsequent calculations. At small inclinations, $\Psi_{2,2} \approx \Psi_{2,2,0}$, an equality which is exactly true at $i = 0^\circ$, since $d_{\mu,m}^l\left(0\right) = \delta_{\mu,m}$. In other words, at small inclinations, the $\mu = 2$ component of the binary potential is primarily responsible for exciting the $m = 2$ disturbances in the disc. Also note that this is the only non-zero component when $e = 0$, since the others have $|n| > 0$, and $\Psi_{m,\mu,n} \propto e^{|n|}$. At intermediate inclinations, the azimuthal components of the binary orbit other than $\mu = 2$ become important, especially the $\mu = 0$ term ($\Psi_{2,0,2}$), which contributes strongly between about $30^\circ$ and $150^\circ$. At inclinations close to $180^\circ$, the component $\Psi_{2,-2,4}$ becomes dominant. This component, despite being somewhat suppressed by its eccentricity dependence, becomes significant due to the strong coupling of the $\mu = -2$ and $m = 2$ terms near counter-alignment. Thus, an eccentric counter-aligned binary can exert non-zero Lindblad torques on a circumstellar disc. This is also true for circumbinary discs (see Section \ref{sec:cb_disc}).

The inclination dependence of each component is determined by a weighted sum of Wigner functions, but to leading order is determined by the first term in this sum. For example, $\Psi_{2,2,0} \propto d_{2,2}^2\left(i\right) \propto \cos^4\left(i/2\right)$ and $\Psi_{2,0,2} \propto d_{0,2}^2\left(i\right) \propto \cos^2\left(i/2\right)\sin^2\left(i/2\right)$, both dominated by the $l = 2$ terms. However, the leading order term is not necessarily a good approximation to the actual $i$ dependence. For example, the leading term of $\Psi_{2,2,0}$ is reduced (compared to $i = 0^\circ$) by $1/4$ at $90^\circ$, while including all terms shows that this reduction is actually about $1/10$, a much steeper decrease with inclination than given by $d_{2,2}^2\left(i\right)$. The shape of the total $\Psi_{2,2}$ is therefore a sum of many different Wigner functions, leading to its characteristic shape. There are two unique characteristics of this curve. First, it changes sign twice, at $i = 29^\circ$ and $i = 172^\circ$ (the particular values of these angles depend on the binary eccentricity), so that no torque is exerted on the disc at these inclinations. The sign of $\Psi_{2,2}$ is irrelevant since the torque is proportional to $|\Psi_{2,2}|^2$. Second, properly summing the relevant $\Psi_{m,\mu,n}$'s can counterintuitively lead to a $\Psi_{m,N}$ whose absolute value (and hence resultant torque) is smaller than $\Psi_{m,m,N-m}$ (the only non-zero component for an aligned disc), as it does for a large range of $i$ in Figure \ref{fig:psi}. 

Figure \ref{fig:torques_cs_i} shows how the resonant torques responsible for disc truncation (normalized by viscous torque) vary with inclination, for an equal mass binary with several different eccentricities. Notably, even for a circular binary (the left panel of Figure \ref{fig:psi}), there are non-zero torques with $N > 2$ when the disc is not aligned. For example, $|T_{2,4}|$ depends on $\Psi_{2,4}$, which is equal to $\Psi_{2,4,0}$ for a circular binary. To leading order this depends on $i$ as $d_{4,2}^4\left(i\right) \propto \cos^6\left(i/2\right)\sin^2\left(i/2\right)$, which is largest at $i = 60^\circ$, and zero at $0^\circ$ and $180^\circ$. In our canonical disc model (with $h = 0.05$ and $\alpha = 0.01$), $|T_{2,4}|$ is never strong enough to clear a gap when $e = 0$, but if $\alpha$ or $h$ of the disc were slightly smaller, it would be possible for a certain range of inclinations. This is impossible when the disc and binary are aligned. The largest torque for a circular binary is $|T_{2,2}|$, which depends only on $\Psi_{2,2,0}$, a monotonically decreasing function of $i$ (see Figure \ref{fig:psi}). This torque is very large at small inclination (exceeding $|T_\nu|$ by a factor of over $10^4$), and gets weaker as inclination increases, becoming about $18$ times smaller (compared to aligned) at $90^\circ$, and falling to zero at $180^\circ$.

For $e > 0$ (the middle and right panels of Figure \ref{fig:torques_cs_i}), higher $N$ torques generally become stronger (compared to $e = 0$). At small inclinations, $|T_{2,2}| > |T_{2,3}| > |T_{2,4}|$ and so on. This is because at small $i$ these torques are primarily a result of the $\Psi_{2,2,N-2}$ potential components, which have the approximately the same inclination dependence [$d_{2,2}^2\left(i\right)$ to leading order], but different eccentricity dependence ($e^{N-2}$), the latter of which determines their relative strengths. At larger inclinations, the contributions of the $\mu \neq 2$ potential components cause the torques to have more oscillatory inclination dependences (as demonstrated in Figure \ref{fig:psi}). The dashed lines in the middle and right panels of Figure \ref{fig:torques_cs_i} show $|T_{2,2}|$ computed using only the $\Psi_{2,2,0}$ component, demonstrating the relative importance of these couplings. As a result of this behavior, at large $i$ (e.g., above about $120^\circ$ for $e = 0.3$), the torque $|T_{2,N}|$ is no longer monotonic in $N$.

\subsection{Location of Outer Disc Edge}

\begin{figure*}
\begin{center}
\includegraphics[width=0.667\textwidth,clip]{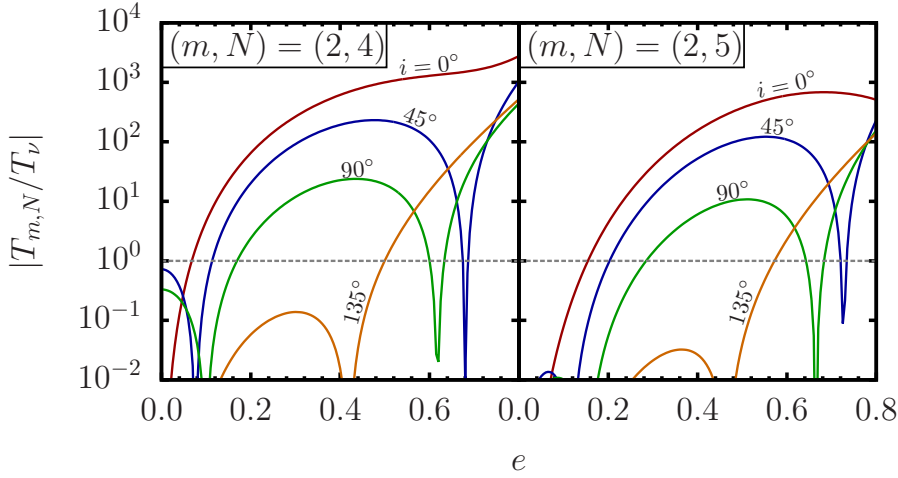}
\caption{Ratio of resonant torque to viscous torque as a function of $e$ for $(m,N) = (2,4)$ and $(2,5)$ ILRs in a circumstellar disc, for several inclinations. The horizontal dashed line corresponds to $|T_{m,N}| = |T_\nu|$.}
\label{fig:torques_cs_e}
\end{center}
\end{figure*}

\begin{figure*}
\begin{center}
\includegraphics[width=0.749\textwidth,clip]{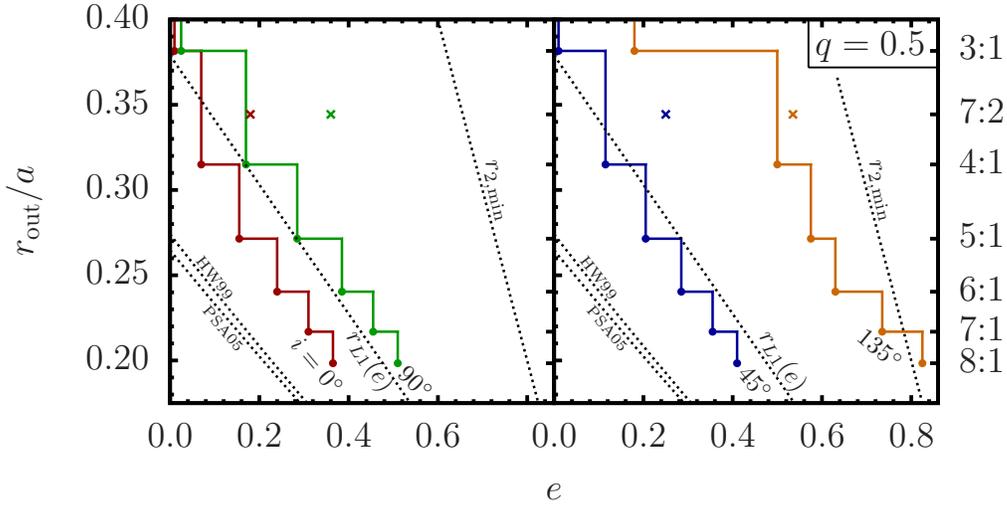}
\caption{Size of a circumstellar disc (location of outer edge $r_\mathrm{out}$) in an equal mass binary as a function of eccentricity for several disc-binary inclinations. The disc parameters are $h = H/r = 0.05$ and $\alpha = 0.01$. The ratios $\Omega(r_{ILR})/\Omega_\mathrm{B}$ for the resonances considered are labeled on the right. The filled points indicate the eccentricities and resonance locations at which the outer disc radius changes, and the lines connecting them indicate the behavior of $r_\mathrm{out}$ between these points. The points represented by crosses indicate the minimum eccentricities required to open a gap at the $\Omega/\Omega_\mathrm{B} = 7$:$2$ commensurability (for inclinations matched to the lines by color). This is an example of a resonance which is not important for disc truncation, since other resonances can open gaps at smaller radii for the same value of $e$. From left to right (in each panel), the four dashed lines correspond to: the size of a particle disc determined by Pichardo, Sparke \& Aguilar (2005) (labeled ``PSA05''), the long-term stability limit for S-type planets in binaries (for $i = 0^\circ$) from Holman \& Wiegert (1999) (labeled ``HW99''), the average Roche lobe radius evaluated at the pericenter separation of the binary [labeled ``$r_{L1}(e)$'', see equation (\ref{eq:eggleton})], and the binary pericenter separation, $r_{2,\mathrm{min}} = (1-e)a$, which sets a strict upper limit for the disc size.}
\label{fig:size_cs_first}
\end{center}
\end{figure*}

The following procedure is used to compute $r_\mathrm{out}$, the location of the outer edge of a circumstellar disc, for a given disc-binary inclination $i$. For each resonance (ILR), labeled by $(m,N)$, or by $\Omega(r_\mathrm{ILR})/\Omega_\mathrm{B} = N/(m-1)$, we first compute the torque $|T_{m,N}|$ as a function of $e$ (see Figure \ref{fig:torques_cs_e}). Then we find the range of $e$ for which $|T_{m,N}| > |T_\nu|$ (so that a gap can be opened), for each $(m,N)$. Then at every value of $e$, the outer radius of the disc is identified as the location of the gap-opening resonance located at the smallest radius.

The torques $|T_{m,N}|$ (normalized by viscous torque at the ILR, $|T_\nu|$) for $(m,N) = (2,4)$ and $(2,5)$ as a function of $e$ are shown in Figure \ref{fig:torques_cs_e}. For small $e$ and $i$, $|T_{m,N}|$ is a monotonically increasing function of $e$, so that finding the range for which a gap can be opened amounts to finding the minimum $e$ for which $|T_{m,N}| > |T_\nu|$. However, for large inclinations, $|T_{m,N}|$ has local extrema which result in the existence of multiple values of $e$ for which $|T_{m,N}| = |T_\nu|$. For example, at $i = 45^\circ$ and $90^\circ$, as $e$ is increased, $|T_{m,N}/T_\nu|$ exceeds unity above a critical $e$, then  reaches a maximum, decreases, and again drops below unity at a second critical $e$. Another minimum is then reached, and then it exceeds unity again at yet another critical $e$. Despite the oscillatory features in Figure \ref{fig:torques_cs_e}, once the $(m,N)$ resonance is cleared, the $(m,N+1)$ resonance is always cleared before $|T_{m,N}|$ becomes non-gap-opening. In other words, as $e$ is increased, the innermost gap clearing resonance shifts inward, so that the disc size is a monotonically decreasing function of $e$ for all inclinations we have considered.

Figure \ref{fig:size_cs_first} shows an example of the disc outer radius as a function of eccentricity for an equal mass binary, for several inclinations. In this figure, the filled points represent the eccentricities and resonant locations at which the outer radius abruptly changes due to a new innermost resonance ``turning on'' (becoming able to clear a gap). These result in the outer disc radius following the stairstep-shaped curves which connect the filled points. Several limiting radii are also shown in Figure \ref{fig:size_cs_first}. The dashed line labeled $r_{L1}(e)$ is the average radius of the Roche lobe around $M_1$ (Eggleton 1983) evaluated at the pericenter of the binary orbit:
\be
\label{eq:eggleton}
r_{L1}(e) = \frac{0.49q_1^{2/3}(1-e)a}{0.6q_1^{2/3}+\ln\left(1+q_1^{1/3}\right)},
\ee
where $q_1 = M_1/M_2 = (1-q)/q$ (with $q = M_2/M_\mathrm{tot}$). Equation (\ref{eq:eggleton}) is approximate since the original Roche lobe is calculated for circular, synchronized binaries. We see from Figure \ref{fig:size_cs_first} that for $i \gtrsim 90^\circ$, our predicted disc radius (based on gap opening criterion) is larger than $r_{L1}(e)$, suggesting that our result for $r_\mathrm{out}$ may be larger than the actual disc size. Note, however, that since equation (\ref{eq:eggleton}) approximates the Roche lobe as a sphere, it is possible for the disc to extend beyond $r_{L1}(e)$ for some inclinations. The dashed line labeled $r_{2,\mathrm{min}}$ corresponds to the pericenter separation of the binary [$r_{2,\mathrm{min}} = (1-e)a$]. This is a strict upper limit on the disc size when it is either aligned or counter-aligned (the disc-projected closest approach distance is slightly different when there is misalignment), since if one of the stars were allowed to plunge through the disc it would rapidly clear material from its orbit.

In Figure \ref{fig:size_cs_first} we have also shown the long-term stability limit for S-type planets in binaries (for $i = 0^\circ$), based on the fitting formula of Holman \& Wiegert (1999). This limit is smaller than the truncation radius of the disc, but close to the ``orbit-crossing'' limit determined by Pichardo, Sparke \& Aguilar (2005).

\begin{figure}
\begin{center}
\includegraphics[width=0.49\textwidth,clip]{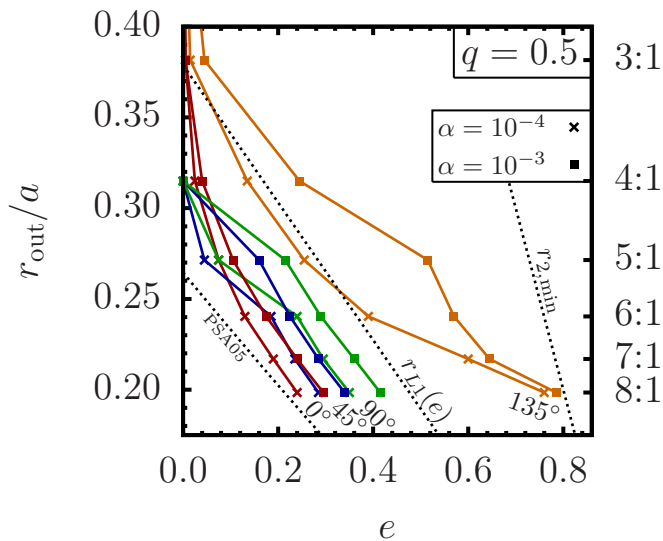}
\caption{Same as Figure \ref{fig:size_cs_first}, except for for discs with viscosity parameter $\alpha = 10^{-3}$ and $10^{-4}$. The straight lines connecting the points are for graphical convenience only, and are a proxy for the starstep-shaped curves depicted in Figure \ref{fig:size_cs_first}.}
\label{fig:size_cs_alpha}
\end{center}
\end{figure}

\begin{figure*}
\begin{center}
\includegraphics[width=0.85\textwidth,clip]{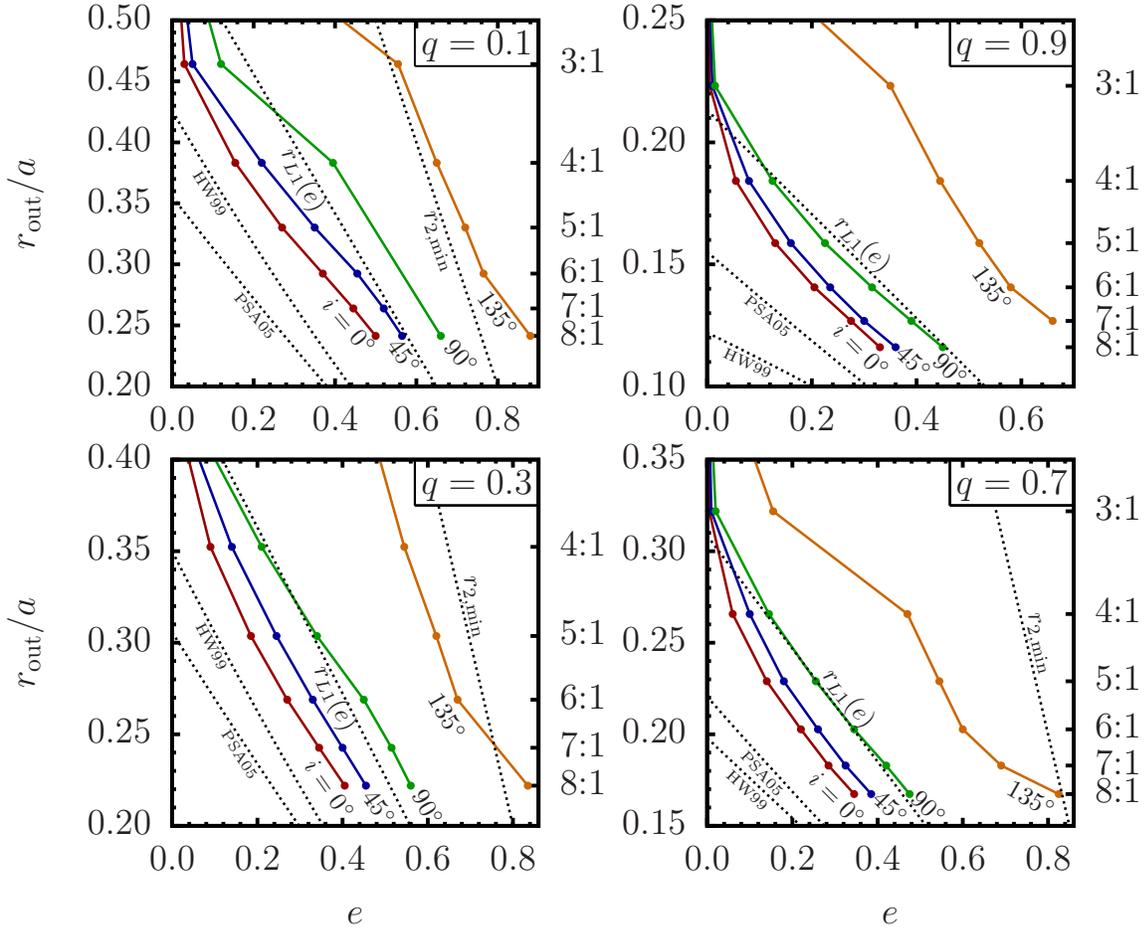}
\caption{Same as Figure \ref{fig:size_cs_first}, except for several different mass ratios. As in Figure \ref{fig:size_cs_alpha}, the straight lines connecting the points are only for graphical convenience, and should be interpreted as stairstep-shaped depicted in Figure \ref{fig:size_cs_first}.}
\label{fig:size_cs_multi}
\end{center}
\end{figure*}

Our canonical disc model has $h = 0.05$ and $\alpha = 0.01$. The effect of varying the disc properties is shown in Figure \ref{fig:size_cs_alpha}, which depicts the disc outer radius as a function of eccentricity for an equal mass binary (as in Figure \ref{fig:size_cs_first}) when $\alpha = 10^{-3}$ and $\alpha = 10^{-4}$. These correspond to a reduction of all viscous torques $T_\nu$ by a factor of $10$ and $100$ compared to our canonical model. Since $T_\nu$ is proportional to $\alpha h^2$, the curves labeled $\alpha = 10^{-3}$ represent any disc with a combination of $\alpha$ and $h$ such that $\left(\alpha/10^{-3}\right)\left(h/0.05\right)^2 = 1$, and similarly for curves labeled $\alpha = 10^{-4}$. Although the viscous torques rescale trivially in this way, the complicated dependence of the resonant torques on eccentricity and inclination does not allow a simple rescaling of the disc size versus eccentricity relations. This explains the qualitative differences between the curves in Figures \ref{fig:size_cs_first} and \ref{fig:size_cs_alpha}. Most notably, for the values of $\alpha$ in Figure \ref{fig:size_cs_alpha}, a gap can be opened at the $4$:$1$ commensurability at zero eccentricity for $i = 45^\circ$ and $90^\circ$, contrary to our canonical disc model (see Figure \ref{fig:size_cs_first}). However, broadly speaking, the difference in disc size due to an order of magnitude change in $\alpha h^2$ is comparable to the difference in size due to a change of inclination of about $45^\circ$.

The effect of binary mass ratio on circumstellar disc size is explored in Figure \ref{fig:size_cs_multi}. Recall our definition of the mass ratio $q = M_2/M_\mathrm{tot}$, and that $M_2$ is always considered the perturber, regardless of whether it is more or less massive than $M_1$. Therefore the disc sizes for $q = 0.1$ and $q = 0.9$ can be thought of as the sizes of the circumprimary and circumsecondary discs in a system in which the secondary is $1/9$ the mass of the primary, and similarly for $q = 0.3$ and $q = 0.7$ (in which case the secondary is $3/7$ the mass of the primary). The behavior is qualitatively similar to the equal mass case (see Figure \ref{fig:size_cs_first}), but with more massive perturbers leading to smaller discs and vice versa.

We have restricted our calculation of $r_\mathrm{out}$ to binaries with $e \lesssim 0.8$. At high eccentricity, our ability to compute the disc outer radius is restricted by our finite expansion of the disturbing potential: the highest order resonance we consider is $\Omega/\Omega_\mathrm{B} = 8$:$1$. For larger values of $e$, higher order resonances can clear gaps, further reducing the size of the disc, but computing the appropriate potential components for these resonances is impractical in our semi-analytic approach. 

\section{Results: Circumbinary Disc}
\label{sec:cb_disc}

\subsection{Resonances Relevant to Inner Disc Truncation}

Determining the size $r_\mathrm{in}$ of the inner cavity of a circumbinary disc amounts to finding the largest radius at which a gap can be cleared. Therefore we consider the OLRs of the potential components with the smallest possible pattern frequencies. Since $\Omega(r_\mathrm{OLR})/\Omega_\mathrm{B} = N/(m+1)$ [see equation (\ref{eq:lr_omega})], we shall focus on $N = 1$ and $m \geq 1$, corresponding to the $\Omega/\Omega_\mathrm{B} = 1$:$(m+1)$ commensurabilities. However, for equal-mass or nearly-equal mass binaries, the strength of the odd-$m$ potential components is zero or small relative to the even-$m$ components, resulting in a relatively weak torque $|T_{m,N}|$. As we will also show, $|T_{1,1}|$ can be weak even for binaries which are not close to equal mass. Therefore, it is also necessary to consider the next strong resonance interior to the $1$:$2$ commensurability, namely the $2$:$3$ commensurability ($m = 2$, $N = 2$), which can be cleared by the $T_{2,2}$ torque. 

\subsection{Effects of Disc Inclination}

\begin{figure*}
\begin{center}
\includegraphics[width=0.99\textwidth,clip]{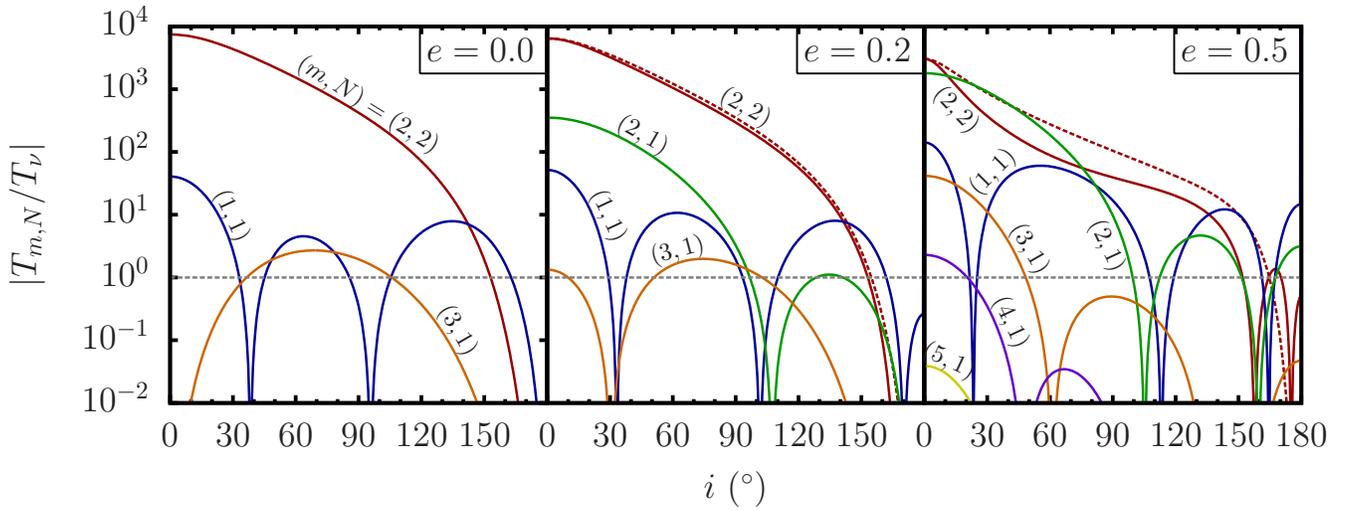}
\caption{Ratio of resonant torque to viscous torque ratio as a function of inclination for the $(m,N) = (m,1)$ and $(2,2)$ outer Lindblad resonances in a circumbinary disc with mass ratio $q = 0.3$. Each panel corresponds to a different binary eccentricity. As in Figure \ref{fig:torques_cs_i}, the horizontal dashed line corresponds to $|T_{m,N}| = |T_\nu|$ (above which a resonance can clear a gap), and the dashed lines in the middle and right panels are the $(2,2)$ torques computed using only the $\Psi_{2,2,0}$ potential component.}
\label{fig:torques_cb_i}
\end{center}
\end{figure*}

\begin{figure*}
\begin{center}
\includegraphics[width=0.667\textwidth,clip]{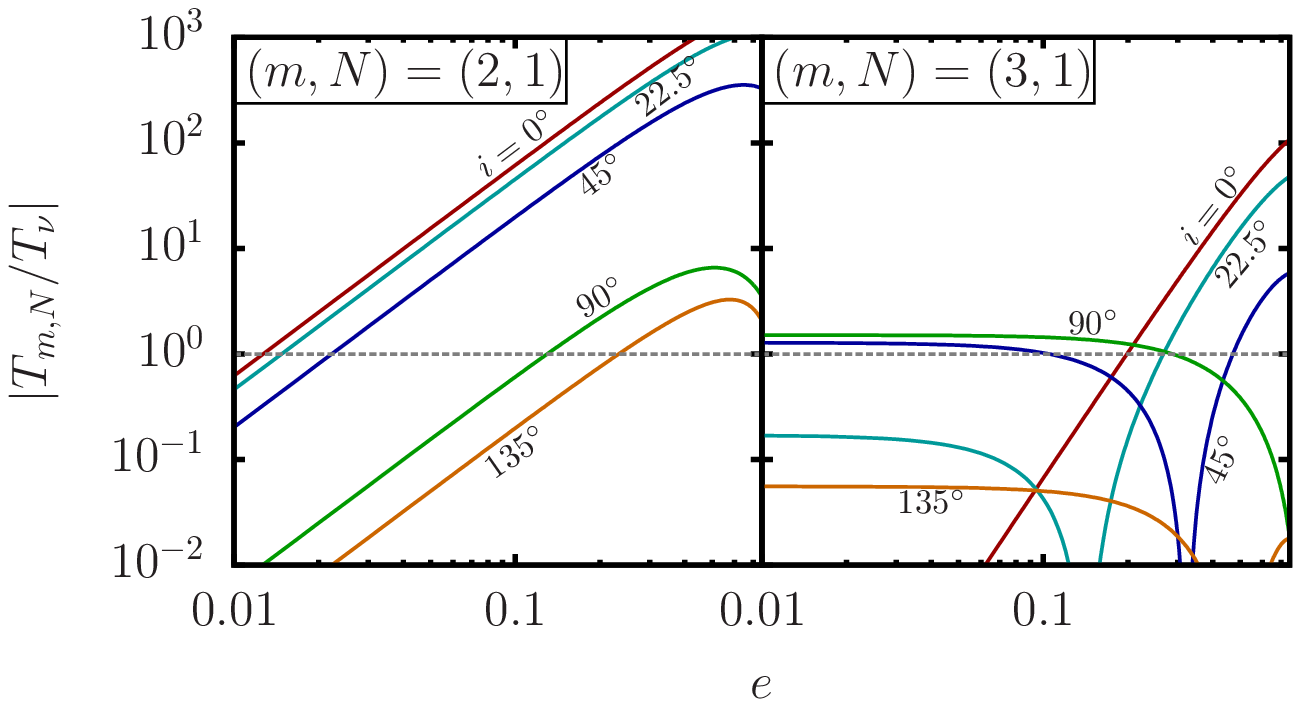}
\caption{Ratio of resonant torque to viscous torque as a function of $e$ for $(m,N) = (2,1)$ and $(3,1)$ outer Lindblad resonances in a circumbinary disc, for a mass ratio $q = 0.3$. The dashed horizontal line indicates $|T_{m,N}| = |T_\nu|$.}
\label{fig:torques_cb_e}
\end{center}
\end{figure*}

Figure \ref{fig:torques_cb_i} shows the resonant torques normalized by viscous torque for a circumbinary disc around a binary with mass ratio $q = 0.3$ (cf. Figure \ref{fig:torques_cs_i} for a circumstellar disc). The $(m,N) = (1,1)$ torque is weak due to the absence of the dipole term in the binary potential [see equation (\ref{eq:cb_potential})], so that to leading order, $\Psi_{1,1}$ is proportional to $d_{1,1}^3\left(i\right) \left(r/a\right)^{-4}$. Meanwhile $\Psi_{2,2}$ is approximately proportional to $d_{2,2}^2\left(i\right)\left(r/a\right)^{-3}$ (and has its OLR at a smaller radius), and so the $(m,N) = (2,2)$ torque is generally much stronger than the $(1,1)$ torque and can potentially be the most relevant to clearing the inner cavity at low $e$ and $i$. The $(2,2)$ component is also relevant for an equal-mass binary, for which all torques with odd $m$ are zero.

For the same reasons as for a circumstellar disc, the torque can be an oscillatory function of $i$, but this effect is even more pronounced for a circumbinary disc. For example, for a circular binary, $T_{3,1}$ can clear a gap at the $1$:$4$ commensurability for $i$ between $37^\circ$ to $106^\circ$, which is impossible for an aligned disc. For $e > 0$, there are several ranges of $i$ for which the ordering of torques is very different than for an aligned disc, for which the torque strengths depend monotonically on the resonance location (innermost resonances are the strongest).

\subsection{Inner Cavity Size: Inner Disc Radius}

\begin{figure*}[p]
\begin{center}
\includegraphics[width=0.99\textwidth,clip]{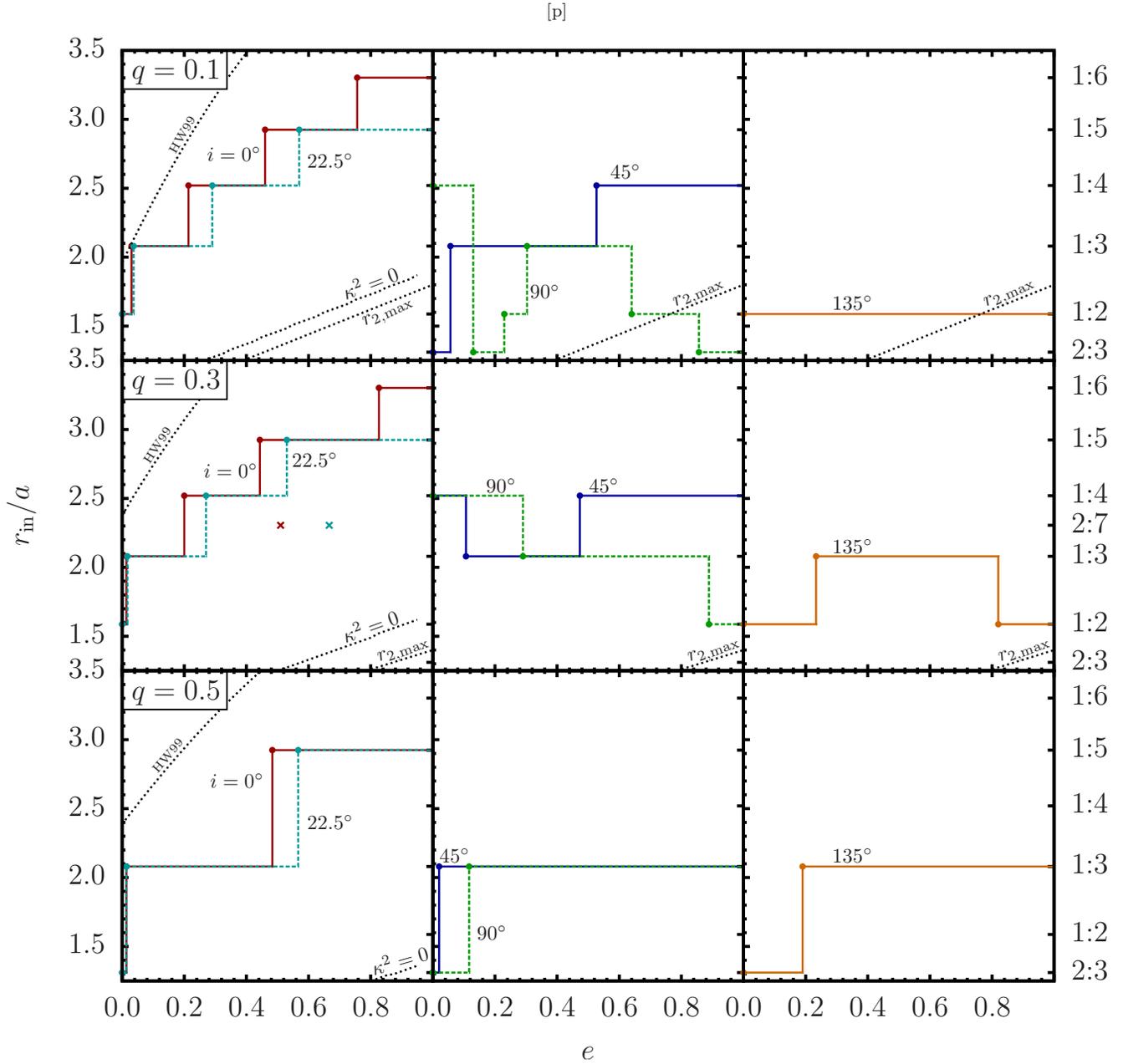}
\caption{Size of inner cavity (location of inner edge $r_\mathrm{in}$) of a disc around a binary with several mass ratios (corresponding to different rows), each for several binary-disc inclinations. The disc parameters are $h = 0.05$ and $\alpha = 0.01$. The ratios $\Omega(r_\mathrm{OLR})/\Omega_\mathrm{B}$ for the resonances considered are labeled on the right. The filled points indicate the eccentricities and resonance locations at which the inner disc radius changes, and the lines connecting them indicate the behavior between these points. In the middle row, the points represented by crosses indicate the minimum eccentricities for which a gap can be cleared at the $2$:$7$ commensurability, for $i = 0^\circ$ and $22.5^\circ$ (for higher inclinations, a gap cannot be cleared for any value of $e$). This is an example of a resonance which is not important in determining the size of the inner disc, since gaps can be opened at larger radii for the same values of $e$. The dashed lines in the left panels labeled ``$\kappa^2 = 0$'' indicate the radius below which the disc does not satisfy the Rayleigh stability criterion (for $i = 0^\circ$). The dashed lines labeled ``$r_{2,\mathrm{max}}$'' correspond to the radial location of the secondary component of the binary at apocenter [$r_{2,\mathrm{max}} = (1-q)(1+e)a$], which is a strict lower limit on $r_\mathrm{in}$ (for $q = 0.5$, $r_{2,\mathrm{max}}$ is below the displayed range of the $y$-axis). The dashed lines labeled ``HW99'' indicate the stability limit for P-type planets in binaries (for $i = 0^\circ$) from Holman \& Wiegert (1999). Note that for $q = 0.3$ and $q = 0.5$, when $i = 0^\circ$ and $i = 22.5^\circ$, the cavity transitions from its smallest size (the $1$:$2$ commensurability for $q = 0.3$ and the $2$:$3$ commensurability for $q = 0.5$) to a larger size ($1$:$3$ in both cases) at a very small but non-zero eccentricity ($e = 0.01 - 0.02$).}
\label{fig:size_cb}
\end{center}
\end{figure*}

The procedure for computing the inner disc radius $r_\mathrm{in}$ for a circumbinary disc at a given inclination $i$ is as follows. First, for each resonance $(m,N) = (m,1)$ and $(2,2)$, we compute $|T_{m,N}|$ as a function of $e$ (see Figure \ref{fig:torques_cb_e}), and determine the range of $e$ values where gap opening is possible ($|T_{m,N}| \geq |T_\nu|$). The inner disc radius is determined by which gap-opening resonance is located at the largest radius for every value of $e$. 

As in the circumstellar disc case, the complicated dependence of $|T_{m,N}|$ on $e$ can result in several ranges of $e$ for which a resonance can open a gap. But unlike the circumstellar disc case, this effect is significant enough to qualitatively alter the dependence of the disc size on $e$. As an example, $|T_{m,N}/T_\nu|$ is shown in Figure \ref{fig:torques_cb_e} for $(m,N) = (2,1)$ and $(3,1)$, corresponding to the $\Omega/\Omega_\mathrm{B} = 1$:$3$ and $1$:$4$ commensurabilities, for a binary with mass ratio $q = 0.3$, for several disc inclinations. While $|T_{2,1}|$ is a monotonically increasing function of $e$, such that for each $i$ there is a critical $e$ above which it can open a gap, the behavior of $|T_{3,1}|$ is very different. For $i = 45^\circ$ and $90^\circ$, it can open a gap at zero eccentricity, and it decreases with $e$ so that above a certain critical $e$ a gap can no longer be opened (for $i = 45^\circ$ a gap can again be opened at yet another larger $e$). In general, a given resonance, $(m,N)$, is not always gap-opening at the lowest value of $e$ for which the next resonance, $(m+1,N)$, first opens a gap. This is in contrast to the case of a circumstellar disc, for which gaps are always cleared at sequentially higher order resonances as $e$ is increased. At large inclinations, this vastly different behavior of the $|T_{m,N}|$'s can result in cavity sizes which both increase or decrease with increasing $e$, sometimes exhibiting both behaviors for a single inclination.

The inner disc radii for three different mass ratios and various inclinations are shown in Figure \ref{fig:size_cb}. We first focus on the middle row, which corresponds to the mass ratio $q = 0.3$ (as in Figures \ref{fig:torques_cb_i} and \ref{fig:torques_cb_e}). For $i = 0^\circ$ and $22.5^\circ$, the inner edge is located at the $1$:$2$ commensurability for $e = 0$ and increases with $e$ as gaps are opened at higher-order resonances. The maximum cavity size is located at the $1$:$6$ commensurability $i = 0^\circ$, and at the $1$:$5$ commensurability for $i = 22.5^\circ$. For $i = 45^\circ$ and $90^\circ$, the $1$:$4$ commensurability can be cleared at zero eccentricity (see the left panel of Figure \ref{fig:torques_cb_i}), but the cavity size then decreases with $e$ as the $1$:$4$ torque becomes too small and the $1$:$3$ torque becomes responsible for gap opening. For $i = 45^\circ$, the size then goes back up to the $1$:$4$ commensurability at $e = 0.47$ and remains at that size for larger $e$. For $i = 90^\circ$ the cavity shrinks as $e$ increases, then remains at the $1$:$2$ commensurability. Clearly, the behavior of the disc inner cavity size is more complicated than in the circumstellar disc case, and even its qualitative behavior (increasing or decreasing with $e$) cannot be trivially ascertained, but requires a full consideration of the details of the resonant torques.

The top and bottom rows of Figure \ref{fig:size_cb} give the inner disc radii for $q = 0.1$ and $q = 0.5$. Note that a mass ratio $q$ greater than $1/2$ is degenerate with a mass ratio of $1-q$ for a circumbinary disc. Much of the qualitative behavior seen in the $q = 0.3$ case can also be seen for $q = 0.1$, for example the cavity size is an increasing function of $e$ for $i \leq 45^\circ$, while for $i = 90^\circ$, it can either be decreasing or increasing over different ranges of $e$. A new behavior is seen for $i = 135^\circ$ and $q = 0.1$: the disc size is indepedendent of $e$. In this case, for all values of $e$, the torque $|T_{1,1}|$ is strong enough to clear the $1$:$2$ commensurability, but no other torque $|T_{m,N}|$ can clear a gap at its resonant location in the disc regardless of the value of $e$. For an equal mass binary ($q = 0.5$), there are fewer possible truncation sites because odd $m$ torques are zero (since $M_l = 0$ for odd $l$). The only allowed cavity sizes are at the $2$:$3$, $1$:$3$ and $1$:$5$ commensurabilities. The dependence of the cavity size on $e$ is simple in this case: for all inclinations, $r_\mathrm{in}$ increases with $e$, with the maximum cavity size located at the $1$:$5$ commensurability for $i = 0^\circ$ and $i = 22.5^\circ$ and at $1$:$3$ commensurability for other inclinations ($45^\circ$, $90^\circ$, $135^\circ$).

Figure \ref{fig:size_cb} also shows, as a function of eccentricity, the radius at which the squared radial epicyclic frequency [see equation (\ref{eq:kappa_non_kep})] is equal to zero. Below this radius, disc particles are unstable to radial perturbations (according to the Rayleigh stability criterion), and so the disc can not extend inwards beyond this radius. This radius is shown in the left panels, and is evaluated for $i = 0^\circ$ (or equivalently, $i = 180^\circ$). It is not shown for larger inclinations, for which the instability region shifts inward and lies interior to $r_{2,\mathrm{max}} = (1-q)(1+e)a$, the radial coordinate of the secondary star at the apocenter of the binary orbit (also shown in Figure \ref{fig:size_cb}), which is a strict lower limit for $r_\mathrm{in}$. For the mass ratios and inclinations we have considered, neither of these non-resonant truncation mechanisms constrain the inner cavity size, except for at very large $e$ for $q = 0.1$ and $i \geq 90^\circ$ (see the two upper right panels), for which the Lindblad torques are weak. Note that for a retrograde disc ($i = 180^\circ$), for which the torques are further weakened, the Rayleigh stability criterion may be more relevant for truncation than gap clearing by Lindblad torques for a wider range of binary parameters (see Nixon \& Lubow 2015).

The long-term stability limit for P-type planets in binaries (for i = $0^\circ$) (Holman \& Wiegert 1999) is also shown in Figure \ref{fig:size_cb}. The inner truncation radius is inside the stability limit, which is of interest to the formation of observed circumbinary planets, many of which reside very close to the stability limit (see Welsh et al. 2014 and references therein). Thus, planets that have formed near or migrated to the inner edge of the disc (e.g., Kley \& Haghighipour 2014) may experience dynamical instability as the disc disappears.

\section{Effects of Non-Keplerian Rotation}
\label{sec:non_kep}

Up to this point, we have considered discs with Keplerian rotation profiles ($\Omega \propto r^{-3/2}$), for which the rotation frequency $\Omega(r)$ is equal to the radial epicyclic frequency $\kappa(r)$. Strictly, this is valid when the static, axisymmetric part of the gravitational potential consists only of a monopole term, proportional to $1/r$. However, in general, the rotation profile is modified by the quadrupole and higher-order components of the binary potential.\footnote{Nixon \& Lubow (2015) included this effect in their analysis of the Lindblad torques experienced by a retrograde circumbinary disc.} The squared rotation frequency is given by
\be
\label{eq:omega_non_kep}
\Omega^2(r) = \frac{1}{r}\frac{\mathrm{d}}{\mathrm{d}r}\left[\Phi_\mathrm{K}(r) + \Phi_{0,0}(r)\right],
\ee
and the squared radial epicyclic frequency by
\be
\label{eq:kappa_non_kep}
\kappa^2(r) = \left(\frac{\mathrm{d}^2}{\mathrm{d}r^2} + \frac{3}{r}\frac{\mathrm{d}}{\mathrm{d}r}\right)\left[\Phi_\mathrm{K}(r) + \Phi_{0,0}\left(r\right)\right].
\ee
Here $\Phi_\mathrm{K}(r)$ is equal to $-GM_1/r$ for a circumstellar disc and $-GM_\mathrm{tot}/r$ for a circumbinary disc, and $\Phi_{0,0}(r)$ is the $(m,N) = (0,0)$ component of the appropriate disturbing potential. Equations (\ref{eq:omega_non_kep}) and (\ref{eq:kappa_non_kep}) assume that pressure gradients, which are an additional source of non-Keplerian rotation, contribute negligibly compared to gravity (see Section \ref{sec:discussion}).

The non-Keplerian rotation profile has two effects on the Lindblad resonances. First, the resonance locations are shifted, so that they no longer correspond to exact integer commensurabilities of $\Omega(r)$ and $\Omega_\mathrm{B}$ [as in equation (\ref{eq:lr_locations})]. Instead, they are determined by solving the general LR condition [equation (\ref{eq:lr_criterion})], and are functions of $e$ and $i$. We adopt the notation $r_\mathrm{LR}^\mathrm{K}$ for the Keplerian LR locations [as given by equation (\ref{eq:lr_locations})] to distinguish them from exact locations, $r_\mathrm{LR}$. Second, the Lindblad torques [equation (\ref{eq:lr_torque})] are modified from their values in a Keplerian disc, since $\mathrm{d}D/\mathrm{d}\ln r$ and $\Psi_{m,N}$ must be evaluated at the new resonance locations using the modified expressions for $\Omega$ and $\kappa$. We also adopt the notation $T_{m,N}^\mathrm{K}$ for the torque on a Keplerian disc to distinguish it from the true torque $T_{m,N}$. In this section, we consider how these effects modify the properties of the resonances relevant to truncation for both circumstellar and circumbinary discs.

\begin{figure*}
\begin{center}
\includegraphics[width=0.99\textwidth,clip]{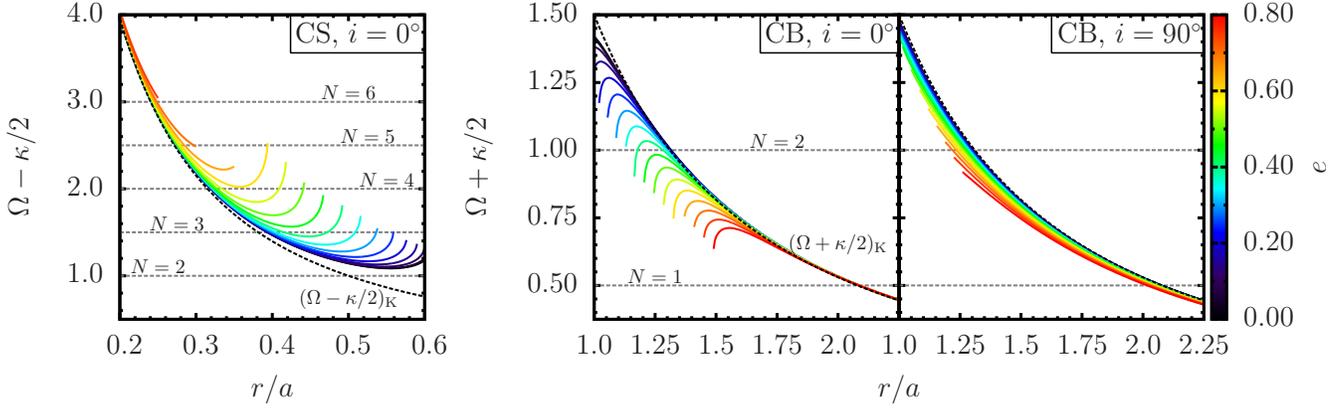}
\caption{The resonant frequencies (combinations of rotation frequency $\Omega$ and radial epicyclic frequency $\kappa$) which determine the location of several Lindblad resonances [see equation (\ref{eq:lr_criterion})]: $\Omega - \kappa/2$ (in units of the binary orbital frequency $\Omega_\mathrm{B}$), which is relevant to $m = 2$ ILRs for a circumstellar disc (shown for an equal mass binary with $i = 0^\circ$; left), and $\Omega + \kappa/2$, which is relevant to $m = 2$ OLRs (shown for a circumbinary disc with $q = 0.3$, for $i = 0^\circ$ and $i = 90^\circ$; right). The line colors correspond to different binary eccentricities, as indicated by the color bar on the right. The curves terminate (at large $r$ for a circumstellar disc, and at small $r$ for a circumbinary disc) at the point at which $\kappa^2$ becomes negative. The dashed curves indicate the corresponding frequencies ($\Omega \pm \kappa/2$) in a Keplerian disc. The horizontal dashed lines represent pattern frequencies $\omega_\mathrm{P} = N\Omega_\mathrm{B}/2$ for several values of $N$. The intersection of the $\Omega \pm \kappa/2$ curves with the horizontal lines correspond to the locations of the $(m,N) = (2,N)$ LRs. Note that for a circumstellar disc, $\Omega$ and $\kappa$ do not strongly deviate from their Keplerian values at large inclinations, therefore only the $i = 0^\circ$ case is shown.}
\label{fig:resonance_non_kep}
\end{center}
\end{figure*}

\subsection{Circumstellar Disc}

\begin{figure*}
\begin{center}
\includegraphics[width=0.667\textwidth,clip]{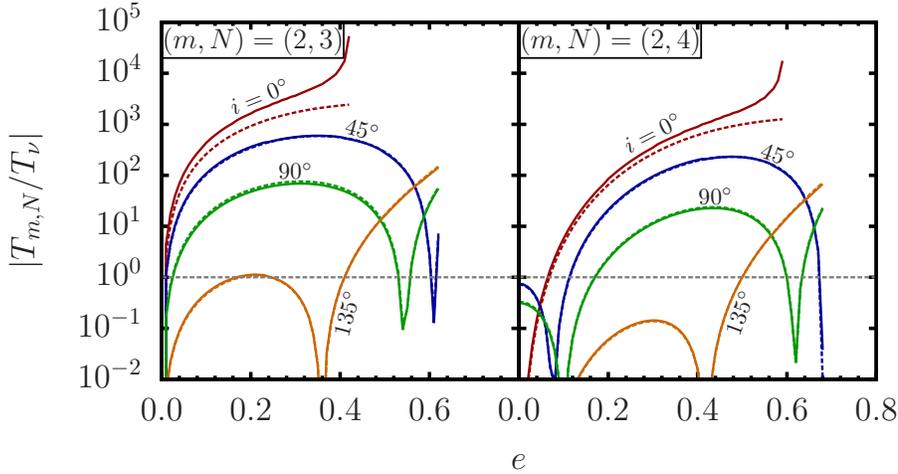}
\caption{Ratio of resonant torque $T_{m,N}$ to viscous torque $T_\nu$, both evaluated at the true resonance location $r_\mathrm{LR}$ (solid lines), and the ratio of these torques in a Keplerian disc, evaluated at the Keplerian resonance location $r_\mathrm{LR}^\mathrm{K}$ (dashed lines), for a circumstellar disc in an equal mass binary. The $(m,N) = (2,3)$ and $(2,4)$ ILRs (left and right panels), which are the lowest-order resonances responsible for truncation at non-zero eccentricity, are each shown for several inclinations. The horizontal dashed line corresponds to the threshold for disc truncation ($|T_{m,N}/T_\nu| = 1$).}
\label{fig:torque_non_kep_cs}
\end{center}
\end{figure*}

Figure \ref{fig:resonance_non_kep} (left panel) gives an example of how the disc rotation profiles and the locations of ILRs (for $m = 2$) are modified by a non-Keplerian potential. The profiles are obtained by evaluating equations (\ref{eq:omega_non_kep}) - (\ref{eq:kappa_non_kep}) numerically. At high eccentricities, the strong deviations from Keplerian rotation render some resonances non-existent, since there is no $r$ for which $\Omega - \kappa/m$ is equal to $\omega_\mathrm{P}$. This occurs for $e \gtrsim 0.4$ for the $N = 3$ ILR, and for $e \gtrsim 0.6$ for the $N = 4$ ILR. Notably, the $(m,N) = (2,2)$ ILR (nominally the $2$:$1$ orbital commensurability) does not exist for an aligned disc in a circular, equal mass binary. The largest fractional shift in $r_\mathrm{LR}$ for a given resonance is about $12\%$, occurring for the largest value of $e$ for which the resonance still exists. For eccentricities near this value, there can also be a second location satisfying the resonant criterion. However, the second location is unlikely to be relevant to disc truncation, as it is close to the point at which the disc becomes Rayleigh unstable, and is located at a larger $r$ than the original resonance.

The numerical results for the ratios of resonant torque $T_{m,N}$ to viscous torque $T_\nu$ are depicted in Figure \ref{fig:torque_non_kep_cs}, along with their Keplerian counterparts. Note that the correction to $T_\nu$ at the modified resonance location has been taken into account in Figure \ref{fig:torque_non_kep_cs}. The deviation of the torque ratio from its Keplerian counterpart is largest for $i = 0^\circ$, and negligible for other inclinations (comparable to the thickness of the plotted lines). The $i = 0^\circ$ curves terminate at a smaller value of $e$ than for other inclinations, due to the fact that the resonance ceases to exist (see Figure \ref{fig:resonance_non_kep}). For eccentricities close to this point, the torque deviates strongly from its Keplerian value (by over an order of magnitue), due to the rotation profile becoming highly distorted. However, these large deviations occur at much larger values of $e$ than the ones for which the resonance first truncates the disc. The eccentricities for which $|T_{m,N}| = |T_\nu|$ change by less than $0.01$ for the resonances shown in Figure \ref{fig:torque_non_kep_cs}. Higher order resonances, located at smaller radii, are affected even less strongly. Therefore, the assumption of a Keplerian disc is a reasonable approximation for determining outer disc truncation.

To quadrupole ($l = 2$) and $e^2$ order, explicit expressions for the disc rotation profiles and Lindblad torques can be obtained. The axisymmetric, time independent part of the disturbing potential can be approximated by
\be
\Phi_{0,0}(r) \approx -\frac{GM_2}{4a}f_\mathrm{CS}(e,i)\left(\frac{r}{a}\right)^2,
\ee
where the dependence on the inclination and eccentricity of the binary is given by
\be
f_\mathrm{CS}(e,i) = \frac{1}{2}\left[3\cos^2(i)-1\right]\left(1+\frac{3}{2}e^2\right).
\ee
Thus, to this order,
\be
\label{eq:omega_kappa_cs}
\Omega^2(r) \approx \Omega^2_\mathrm{K} + \frac{2\Phi_{0,0}(r)}{r^2} \quad \mathrm{and} \quad \kappa^2(r) \approx \Omega^2_\mathrm{K} + \frac{8\Phi_{0,0}(r)}{r^2}.
\ee
The resonance condition for $m = 2$ ILRs, which are relevant to outer disc truncation, is then
\be
\Omega_\mathrm{K}\left(1+2\epsilon_\mathrm{CS}\right) \approx N\Omega_\mathrm{B},
\ee
where we have defined the dimensionless parameter
\be
\label{eq:lr_shift_cs}
\epsilon_\mathrm{CS} = \left(\frac{\Phi_{0,0}}{\Phi_\mathrm{K}}\right)_{r^\mathrm{K}_\mathrm{LR}} = \frac{qf_\mathrm{CS}(e,i)}{4N^2}.
\ee
The resonances are therefore located at
\be
\label{eq:res_shift_cs}
r_\mathrm{LR} \approx r_\mathrm{LR}^\mathrm{K}\left(1 + \frac{4}{3}\epsilon_\mathrm{CS}\right),
\ee
and hence shift outward (inward) compared to their locations in a Keplerian disc when $i$ is less (greater) than $54.7^\circ$. We can use equations (\ref{eq:omega_kappa_cs}) and (\ref{eq:res_shift_cs}) to evaluate the Lindblad torque $T_{m,N}$ [equation (\ref{eq:lr_torque})] explicitly (to quadrupole and $e^2$ order). For $m = 2$, we find
\be
T_{2,N}\left(r_\mathrm{LR}\right) \approx T_{2,N}^\mathrm{K}\left(r_\mathrm{LR}^\mathrm{K}\right)\left(1 + \frac{58}{3}\epsilon_\mathrm{CS}\right).
\ee

\subsection{Circumbinary Disc}

\begin{figure*}
\begin{center}
\includegraphics[width=0.667\textwidth,clip]{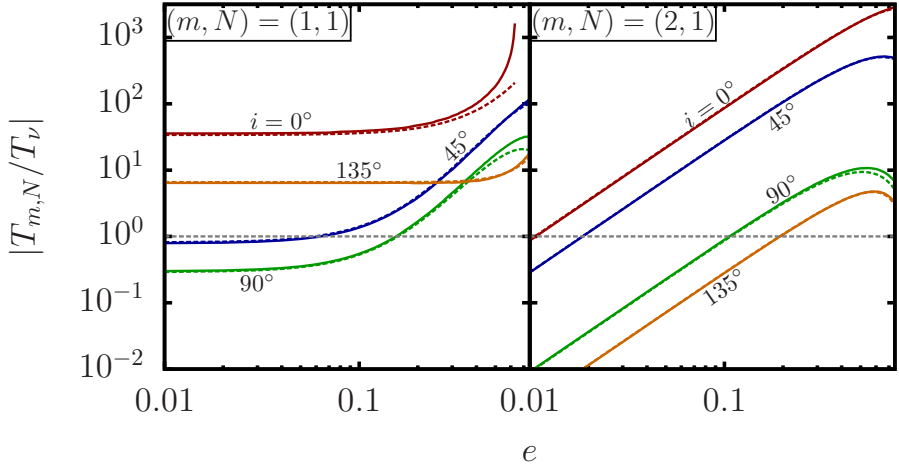}
\caption{Same as Figure \ref{fig:torque_non_kep_cs}, but for the $(m,N) = (1,1)$ and $(2,1)$ OLRs in a circumbinary disc with mass ratio $q = 0.3$.}
\label{fig:torque_non_kep_cb}
\end{center}
\end{figure*}

In Figure \ref{fig:resonance_non_kep} (middle and right panels), we show the locations of the $(m,N) = (2,1)$ and $(2,2)$ OLRs for $i = 0^\circ$ and $90^\circ$, including the effects of the non-Keplerian potential. The largest fractional shifts are about $4\%$, provided that the resonance exists. The $(m,N) = (2,2)$ resonance does not exist for $e \gtrsim 0.5$ when $i = 0^\circ$, and for $e \gtrsim 0.7$ when $i = 90^\circ$.

The ratio of resonant to viscous torques is shown in Figure \ref{fig:torque_non_kep_cb}. The $(m,N) = (1,1)$ torque differs most from the Keplerian case for $i = 0^\circ$, especially near $e \approx 0.6$, above which the resonance no longer exists. For $i = 90^\circ$, there is also some appreciable deviation from the Keplerian torque at large values of $e$. However, the values of $e$ for which the resonances become gap-opening change negligibly. Therefore, as for the case of a circumstellar disc, the approximation of a Keplerian disc does not affect truncation.

To leading order in $l$ and $e$, the non-Keplerian part of the potential can be approximated by
\be
\Phi_{0,0} \approx -\frac{G\mu_\mathrm{B}}{4a}f_\mathrm{CB}(e,i)\left(\frac{r}{a}\right)^{-3},
\ee
with
\be
f_\mathrm{CB}(e,i) = \frac{1}{2}\left\{\left[3\cos^2(i)-1\right]\left(1+\frac{3}{2}e^2\right) - 15e^2\sin^2(i)\right\},
\ee
where $\mu_\mathrm{B} = M_1M_2/M_\mathrm{tot}$ is the reduced mass of the binary. The resulting rotation frequency and epicyclic frequency are
\be
\Omega^2(r) \approx \Omega^2_\mathrm{K} - \frac{3\Phi_{0,0}(r)}{r^2} \quad \mathrm{and} \quad \kappa^2(r) \approx \Omega^2_\mathrm{K} + \frac{3\Phi_{0,0}(r)}{r^2}.
\ee
The resonance condition for $N = 1$ OLRs, which are relevant to inner disc truncation, is
\be
\Omega_\mathrm{K}\left[1+\frac{3}{2}\left(\frac{m-1}{m+1}\right)\epsilon_\mathrm{CB}\right] \approx \frac{\Omega_\mathrm{B}}{m+1},
\ee
where we have defined the dimensionless parameter
\be
\epsilon_\mathrm{CB} = \left(\frac{\Phi_{0,0}}{\Phi_\mathrm{K}}\right)_{r_\mathrm{LR}^\mathrm{K}} = \frac{q(1-q)f_\mathrm{CB}(e,i)}{4(m+1)^{4/3}}.
\ee
So the resonance locations are
\be
r_\mathrm{LR} \approx r_\mathrm{LR}^\mathrm{K}\left[1 + \left(\frac{m-1}{m+1}\right)\epsilon_\mathrm{CB}\right].
\ee
In this approximation, the $(m,N) = (1,1)$ resonance (nominally the $\Omega/\Omega_\mathrm{B} = 1$:$2$ commensurability) is not shifted relative to its position in a Keplerian disc. The other $N = 1$ resonances shift to larger (smaller) radii compared to their locations in a Keplerian disc when $f_\mathrm{CB}(e,i)$ is positive (negative). Explicitly evaluating the Lindblad torque to the same order of approximation, we find that for $N = 1$ (and $m > 1$),
\be
\begin{aligned}
T_{m,1}\left(r_\mathrm{LR}\right) &\approx T_{m,1}^\mathrm{(K)}\left(r_\mathrm{LR}^\mathrm{K}\right) \\ 
&\times \left\{1 + \left[\frac{m+2}{m+1} + \frac{12m}{3m+1} - 2(m-1)\right] \epsilon_\mathrm{CB}\right\}.
\end{aligned}
\ee

\section{Summary and Discussion}
\label{sec:discussion}

\subsection{Main Results}

We have developed a method for computing Lindblad torques due to a binary potential on circumstellar and circumbinary discs which are misaligned with the binary orbital plane. We used this theory to determine the outer radii of circumstellar discs and the inner radii of circumbinary discs, generalizing the work of Artymowicz \& Lubow (1994; AL94) for aligned discs. The summary of our results is as follows. 

In the presence of misalignment (and non-zero eccentricity), each azimuthal component of the binary potential experienced by the disc is a result of many azimuthal components of the potential in the binary plane. This is in contrast to aligned disc, for which each azimuthal component of the potential can only produce perturbations in the disc which have the same azimuthal number. As a result, the inclination and eccentricity dependence of the gap-opening Lindblad torques can be somewhat complicated. Rather than simply decreasing with inclination $i$ (as for circular binaries), or increasing with eccentricity $e$ (as for aligned discs), the Lindblad torque, $|T_{m,N}|$, associated with the potential component which rotates with pattern frequency $\omega_\mathrm{P} = N\Omega_\mathrm{B}/m$ (where $\Omega_\mathrm{B}$ is the binary orbital frequency, and $m$, $N$ are integers), generally has multiple extrema as functions of $e$ and $i$. 

As in AL94, we adopt the resonance gap opening criterion $|T_{m,N}| > |T_\nu|$, where $T_\nu$ is the viscous torque evaluated at the resonance location. For circumstellar discs, the most important resonances for disc truncation are located at the $\Omega$/$\Omega_\mathrm{B} = N$:$1$ commensurabilities, with $N \geq 2$. Ignoring non-resonant truncation mechanisms, the outer edge of the disc is determined by the innermost of these resonances which can clear a gap. Despite the complicated dependence of inner Lindblad torques $|T_{m,N}|$ on inclination and eccentricity, for a given inclination, the resultant outer disc radius is a decreasing function of eccentricity (see Figures \ref{fig:size_cs_first}--\ref{fig:size_cs_multi}). This is the same qualitative behavior as for an aligned disc (AL94). Larger inclinations lead to larger discs: in an equal mass binary, the disc is about $20\%$ larger for $i = 90^\circ$ and $40\%$ larger for $i = 135^\circ$, compared to an aligned ($i = 0^\circ$) disc. 

If the innermost gap-opening resonance lies outside of the tidal radius, then the disc is truncated non-resonantly. This tidal radius has been estimated to be $75 - 90\%$ of the Roche lobe radius for aligned discs (Paczy\'{n}ski 1977; Papaloizou \& Pringle 1977; Paczy\'{n}ski \& Rudak 1980; Pichardo, Sparke \& Aguilar 2005). In the absence of an equivalent theory for misaligned discs, we estimate the tidal radius to be $r_{L1}(e)$, the average Roche lobe radius at the binary pericenter separation [see equation (\ref{eq:eggleton})]. We find that a circumstellar disc in an equal mass binary fills its Roche lobe if inclined more than $45^\circ$, for our standard disc parameters ($h = 0.05$ and $\alpha = 0.01$). If $\alpha h^2$ is smaller by a factor of $10^1 - 10^2$ (a reasonable range for protoplanetary discs), resonant truncation can restrict the disc size to less than $r_{L1}(e)$ for inclinations up to $90^\circ$. In non-equal mass binaries (we have considered $M_2/M_1 = 2.3 - 9.0$), for our standard disc parameters, the circumprimary disc fills its Roche lobe if inclined by more than $45^\circ$, while the circumsecondary disc barely fills its Roche lobe at an inclination of $90^\circ$. Since the Roche lobe formula [equation (\ref{eq:eggleton})] is approximate, 3D numerical calculations are needed to determine the precise outer radii of such discs.

We have also considered the resonance clearing of the inner cavity of a circumbinary disc. The most important resonances are the $\Omega/\Omega_\mathrm{B} = 1$:$N$ ($N \geq 2$) and $2$:$3$ commensurabilities. As for a circumstellar disc, the outer Lindblad torques have a complicated dependence on inclination and eccentricity. However, unlike for a circumstellar disc, the dependence of the inner disc truncation radius on eccentricity can be very different at large inclinations (see Figure \ref{fig:size_cb}). For an aligned disc, the inner cavity radius increases with eccentricity, and cannot extend past the $1$:$6$ commensurability. This is also true at small inclinations (e.g., $22.5^\circ$), for which the cavity is slightly smaller than for an aligned disc. At larger inclinations, the cavity size can either be an increasing or decreasing function of eccentricity, possibly exhibiting both behaviors of different ranges of $e$. Nonetheless, the inner disc radius is generally smaller at large inclinations, for example, never extending past the $1$:$4$ commensurability for $i = 45^\circ$ or $i = 90^\circ$, or past the $1$:$3$ commensurability for $i = 135^\circ$.

\subsection{Approximations and Uncertainties}

The calculations presented in this paper adopt several assumptions. We have assumed the disc to be flat. This is reasonable, as under typical conditions, bending waves and viscous stress lead to small disc warps (see Foucart \& Lai 2014). Our results for the disc truncation radii (for both circumstellar and circumbinary discs) are based on the assumption of exactly Keplerian rotation profiles ($\Omega \propto r^{-3/2}$). The validity of this approximation is addressed in Section \ref{sec:non_kep}. We find that, although non-Keplerian effects due to the binary potential can modify the resonance properties significantly at high eccentricities (e.g., some resonances can be rendered non-existent), gap-clearing by the relevant resonances occurs at low eccentricities and is only slightly modified. Thus, non-Keplerian rotation effects have a negligible effect on disc truncation. We note, however, that we did not consider the effect of strong pressure gradients near the edge of the disc, which may cause further modification of the rotation profile and LRs (Petrovich \& Rafikov 2012).

Our most important assumption (which is also made in AL94) is that angular momentum is deposited into the bulk disc material at, or very close to, the Lindblad resonances. This may affect our results in an appreciable way. First, angular momentum cannot be directly imparted to the disc material at Lindblad resonances, but must be carried as waves (Goldreich \& Nicholson 1989), and received by the disc where the waves dissipate, either due to viscous damping (Takeuchi, Miyama \& Lin 1996) or due to wave steepening and shock formation, although the latter should occur almost immediately after the waves are excited for the mass ratios we have considered (Goodman \& Rafikov 2001). Second, the excited waves have broad angular momentum flux profiles, with widths comparable to their radial wavelengths (e.g., Meyer-Vernet \& Sicardy 1987), rather than being sharply peaked at the Lindblad resonances. For these reasons, the locations of disc truncation computed in this work may differ from those produced by self-consistent numerical treatments which account for these effects. For example, the hydrodynamic simulations of discs aligned with equal mass circular binaries by MacFadyen \& Milosavljevic (2008) show that while the resonance responsible for clearing the inner cavity ($m = 2$, $N = 2$) is formally located at $r = 1.31a$ for a Keplerian disc, the actual cavity extends to nearly $2a$. This suggests that the details of wave excitation, propagation and breaking (or damping) are important uncertainties in the disc truncation process. With this is mind, our results for the truncation radii as functions of binary eccentricity and inclination should be interpreted as general trends rather than exact, sharply defined boundaries. We emphasize the need for detailed numerical work to fully interpret and assess the results of this paper.

\section*{Acknowledgments}

We thank Diego Mu\~{n}oz for help initiating this project and for valuable discussions. We also thank Karl Stapelfeldt for motivating this project, Roman Rafikov for useful discussion, and Stephen Lubow for useful comments. This work has been supported in part by NSF grant AST-1211061, and NASA grants NNX14AG94G and NNX14AP31G.

\appendix

\section{Expansion of Potential as Power Series in Eccentricity}
\label{sec:elliptic}
We expand $\left(r_\mathrm{12}/a\right)^{-\left(l+1\right)} \cos\left(m\phi-\mu \phi'_2\right)$ [for a circumstellar disc, see equation (\ref{eq:cs_expansion})], or $\left(r_\mathrm{12}/a\right)^l \cos\left(m\phi-\mu \phi'_2\right)$ [for a circumbinary disc, see equation (\ref{eq:cb_expansion})], as a power series in $e$. Here $r_{12}$ is the binary separation and $\phi'_2$ specifies the azimuthal position of $M_2$ relative to $M_1$ in the orbital plane of the binary, which, without loss of generality, can be chosen to be the true anomaly of the binary. We use the elliptic expansions (e.g., Murray \& Dermott 1999; Brouwer \& Clemence 1961) of $r_{12}/a$ and $\phi'_2$, in terms of $e$ and mean anomaly $M = \Omega_\mathrm{B}t$. For the binary separation $r_{12}$ we have
\be
\begin{aligned}
\frac{r_{12}}{a} & = 1 + \frac{1}{2}e^2 -2e\sum_{s=1}^\infty \frac{1}{s^2} \frac{\mathrm{d}}{\mathrm{d}e} J_s\left(se\right) \cos\left(sM\right) \\
& = 1 - e\cos\left(M\right) + \frac{e^2}{2}\left[1-\cos\left(2M\right)\right] + \frac{3e^3}{8}\left[\cos\left(M\right)-\cos\left(3M\right)\right] \\
&+ \frac{e^4}{3}\left[\cos\left(2M\right)-\cos\left(4M\right)\right] + \mathcal{O}\left(e^5\right).
\end{aligned}
\ee
The following expression relating the mean anomaly $M$ and eccentric anomaly $E$ is useful:
\be
\begin{aligned}
E & = M + 2\sum_{s=1}^\infty \frac{1}{s} J_s\left(se\right)\sin\left(sM\right) \\
& = M + e\sin\left(M\right) + \frac{e^2}{2}\sin\left(2M\right) + e^3\left[\frac{3}{8}\sin\left(3M\right)-\frac{1}{8}\sin\left(M\right)\right] \\
&+ e^4\left[\frac{1}{3}\sin\left(4M\right)-\frac{1}{6}\sin\left(2M\right)\right] + \mathcal{O}\left(e^5\right).
\end{aligned}
\ee
The true anomaly is given by
\be
\begin{aligned}
\phi'_2 & = \int_0^{M}\frac{\left(1-e^2\right)^{\frac{1}{2}}}{\left[1-e\cos\left(E\right)\right]^2} \mathrm{d}M \\
& = M + 2e\sin\left(M\right) + \frac{5e^2}{4}\sin\left(2M\right) + e^3\left[\frac{13}{12}\sin\left(3M\right)-\frac{1}{4}\sin\left(M\right)\right] \\
&+ e^4\left[\frac{103}{96}\sin\left(4M\right)-\frac{11}{24}\sin\left(2M\right)\right] + \mathcal{O}\left(e^5\right).
\end{aligned}
\ee
Using the above expressions we can obtain the power series (in $e$) for $\left(r_\mathrm{12}/a\right)^{-\left(l+1\right)} \cos\left(m\phi-\mu \phi'_2\right)$ and $\left(r_\mathrm{12}/a\right)^l \cos\left(m\phi-\mu \phi'_2\right)$, and gathering $\cos\left[m\phi-\left(\mu+n\right)\Omega_\mathrm{B}t\right]$ terms [equations (\ref{eq:cs_expansion}) and (\ref{eq:cb_expansion})], we find the coefficients $C^{\mathrm{CS}}_{l,\mu,n}$ and $C^{\mathrm{CB}}_{l,\mu,n}$. 

Using computer algebra, we have computed $C^{\mathrm{CS}}_{l,\mu,n}$ and $C^{\mathrm{CB}}_{l,\mu,n}$ for $|n| \leq 8$. Each coefficient is proportional, to leading order, to $e^{|n|}$, with higher order terms proportional to $e^{|n|+2}$, $e^{|n|+4}$, and so on. We have included terms up to order $e^{10}$, so that every $C_{l,\mu,n}$ includes at least one higher order correction in eccentricity after its leading order. The leading terms for $|n| \leq 4$ are shown below.
\be
\begin{aligned}
C^{\mathrm{CS}}_{l,\mu,0} &= 1 \\
C^{\mathrm{CS}}_{l,\mu,\pm 1} &= \frac{1}{2}e\left[l\pm2\mu+1\right] \\
C^{\mathrm{CS}}_{l,\mu,\pm 2} &= \frac{1}{8}e^2\left[l^2+(5\pm4\mu)l+4\mu ^2\pm9\mu+4\right] \\
C^{\mathrm{CS}}_{l,\mu,\pm 3} &= \frac{1}{48}e^3\left[l^3+6(2\pm\mu)l^2+\left(12\mu^2\pm45\mu+38\right)l \right. \\
&\left. \pm8\mu^3+42\mu^2\pm65\mu+27\right] \\
C^{\mathrm{CS}}_{l,\mu,\pm 4} &= \frac{1}{384}e^4\left[l^4+(22\pm8\mu)l^3+\left(24\mu^2\pm126\mu+155\right)l^2 \right. \\
&\left. +\left(\pm32\mu^3+240\mu^2\pm558\mu+390\right)l+16\mu^4\pm152\mu^3 \right. \\
&\left. +499\mu^2\pm646\mu+256\right]
\end{aligned}
\ee
\be
\begin{aligned}
C^{\mathrm{CB}}_{l,\mu,0} &= 1 \\
C^{\mathrm{CB}}_{l,\mu,\pm 1} &= \frac{1}{2}e\left[-l \pm 2\mu\right] \\
C^{\mathrm{CB}}_{l,\mu,\pm 2} &= \frac{1}{8} e^2\left[l^2-(3 \pm 4\mu)l+\mu(4\mu \pm 5)\right] \\
C^{\mathrm{CB}}_{l,\mu,\pm 3} &= \frac{1}{48} e^3\left[-l^3+3(3 \pm 2\mu)l^2-\left(12\mu^2 \pm 33\mu+17\right)l \right. \\
&\left. +2\mu\left(\pm4\mu^2+15\mu\pm13\right)\right] \\
C^{\mathrm{CB}}_{l,\mu,\pm 4} &= \frac{1}{384} e^4 \left[l^4-2(9 \pm 4\mu)l^3+\left(24\mu^2 \pm 102 \mu+95\right)l^2 \right. \\
&\left. -2\left(\pm 16\mu^3+96\mu^2 \pm 165\mu+71\right)l \right. \\
&\left. + \mu\left(16 \mu ^3 \pm 120 \mu ^2+283 \mu \pm 206\right)\right]
\end{aligned}
\ee

Note that these coefficients are equivalent to the Hansen coefficients $X^{a,b}_c(e)$ (Murray \& Dermott 1999), with
\be
C^{\mathrm{CS}}_{l,\mu,n} = X^{-(l+1),\mu}_{\mu+n}(e) \quad \mathrm{and} \quad C^{\mathrm{CB}}_{l,\mu,n} = X^{l,\mu}_{\mu+n}(e).
\ee
The Hansen coefficients are defined according to
\be
\left[\frac{r_{12}(t)}{a}\right]^{\lambda} \exp\left[-\mathrm{i}\mu\phi'_2(t)\right] = \sum_N X^{\lambda,\mu}_N \exp\left[-\mathrm{i}N\Omega_\mathrm{B}t\right],
\ee
which can be inverted, giving the following expression:
\be
X^{\lambda,\mu}_N(e) = \frac{1}{\pi} \int_0^\pi \frac{\cos\left\{N\left[E-e\sin\left(E\right)\right]-\mu\phi'_2\right\}}{\left[1-e\cos\left(E\right)\right]^{-\left(\lambda+1\right)}} \mathrm{d}E.
\ee
They can also be computed recursively using the Newcomb operators.

\end{document}